\documentclass[journal]{IEEEtran}

\usepackage{caption2}
\usepackage{verbatim}
\usepackage{algorithm}
\usepackage{algorithmic}
\usepackage{sidecap}
\usepackage{amsmath}

\usepackage{latexsym}
\usepackage{color}
\usepackage{amssymb}
\usepackage{cite}
 \usepackage[dvips]{graphicx}
\usepackage[tight,footnotesize]{subfigure}
\usepackage{sidecap}

\graphicspath{{picture/}}
\makeatletter

\newcommand{\Rmnum}[1]{\expandafter\@slowromancap\romannumeral #1@}
\makeatother

\begin{document}
\title{Link Scheduling for Throughput Maximization in Multihop Wireless Networks Under Physical Interference}

\author{\IEEEauthorblockN{Yaqin Zhou\IEEEauthorrefmark{1},
Xiang-Yang Li\IEEEauthorrefmark{2}\IEEEauthorrefmark{3},
Min Liu\IEEEauthorrefmark{1},
Zhongcheng Li\IEEEauthorrefmark{1},
Xiaohua Xu\IEEEauthorrefmark{4}\\
\IEEEauthorblockA{\IEEEauthorrefmark{1}Institute of Computing Technology, Chinese Academy of Sciences, Beijing, China}\\
\IEEEauthorblockA{\IEEEauthorrefmark{2}Illinois Institute of Technology, Chicago, IL, USA}\\
\IEEEauthorblockA{\IEEEauthorrefmark{3}TNLIST, School of Software, Tsinghua University, Beijing, China}\\
\IEEEauthorblockA{\IEEEauthorrefmark{4}University of Toledo, Toledo, OH, USA}
}}

\maketitle

\begin{abstract}
 We consider the problem of link scheduling for throughput  maximization in  multihop wireless networks. Majority of previous methods are  restricted to graph-based interference models. In this paper we study the link scheduling problem using a more realistic physical interference model. Through some key observations about this model, we develop efficient link scheduling algorithms  by exploiting the intrinsic connections between the physical interference model and the graph-based interference model. For one variant of the problem where each node can dynamically adjust its transmission power, we design a scheduling method with $O(g(E))$ approximation  to the optimal throughput capacity where $g(E)$ denotes link diversity. For the other variant where each node has a fixed but possible different transmission powers for different nodes, we design a method with $O(g(E))$-approximation ratio when the transmission powers of all nodes are within a constant factor of each other, and in general with an approximation ratio of $O(g(E)\log\rho )$ where $\log\rho$ is power diversity.
 We further prove that our algorithm  for fixed transmission power case retains $O(g(E))$ approximation  for  any  length-monotone, sub-linear fixed power  setting. Furthermore, all these approximation factors are independent of network size.
\end{abstract}


\section{Introduction}
Various wireless networks (single-hop or multi-hop), e.g., sensor networks, cellular networks, mesh networks, have been deployed  for a broad range of applications.
Of all these networks, a common fundamental problem is to develop efficient scheduling algorithms that can achieve closely to the optimal throughput capacity.
This problem is difficult  because of various challenging factors, especially,  wireless interference, which constraints the set of links that can transmit simultaneously.

Most of previous algorithm design for the link scheduling problem and its variants
simply models wireless interference through geometric graphs, such as conflict graphs and disk graphs\cite{S:ptas}.
In these graph-based models, interference is pairwise and binary.
A set of links is conflict-free if they are pair-wisely conflict-free. Therefore, the interference effect of a transmitter is local and predefined. Intuitively, these features of graph-based interference models enable the link scheduling problems more tractable. We note that many scheduling problems are NP-hard even under this simplified graph-based conflict models \cite{sharma2006complexity}.

While graph-based interference models help to understand these complex link scheduling problems in wireless networks, they do not capture some key features of real wireless communication because they are just simple approximations of the realistic interference constraints.  A more realistic interference model is the so-called physical interference model\cite{S:phy8}. Under the physical interference model, a transmission is successful if the Signal-to-Interference-plus-Noise Ratio (SINR) constraint is satisfied. That is, the ratio of the desired signal strength and the summed interference from all other concurrent transmissions plus ambient noise exceeds some threshold $\sigma$. This additive and global interference, considering transmission powers, has a significant impact on the capacity of a wireless network. Since these features are not captured in graph-based models, a direct application of algorithms under graph-based models may even suffer arbitrarily bad performance \cite{S:phy7}.

Given a time-slotted wireless system, one class of optimum solution to this link scheduling problem for throughput  maximization is to find a maximum weighted independent set of links (MWISL) under the SINR constraint at every time slot \cite{S:MWM1}. Here weight is the queue length of a link. There are two variants of the MWISL problem under the physical interference model,
whether each node has an adjustable or fixed transmission power.
The variant with adjustable transmission power shall jointly solve the power assignment; the variant with fixed transmission power takes predefined powers as  input of the problem.

Existing works mainly focus on approximation algorithms of MWISL for some special cases of power assignment.
Constant approximation ratios are only available for the linear power assignment in literature, with \cite{S:phy5}\cite{halldosson2012infocom} in centralized implementation and \cite{my1} for distributed implementation.
For other power assignments, there are various logarithmic approximation or poly-logarithmic approximation.
For uniform power assignment, \cite{halldosson2012infocom} achieves poly-logarithmic approximation where the logarithmic factors are the size of network, and  \cite{S:phy9}  attains a logarithmic approximation factor that is the ratio between the maximum weight and minimum weight\cite{S:phy9}.
Through algorithmic reductions from maximum independent set of links (MISL)  problem, \cite{S:phy6} proves the existence of  another logarithmic approximation dependent on the cardinality of a maximum independent set of links for adjustable and fixed transmission power. Finding a MISL under physical interference itself is NP hard, and there are no closed-form expressions for size of MISL. Under some extreme cases, the number may be in the order of network  size.

On the other hand, the throughput-optimal link scheduling problem   can be solved using traditional network stability technique, instead of solving MWISL directly. Recently \cite{Pei2012mobihoc} has combined this technique and randomized technique to get $O(g(E))$-fractional capacity region for a special case of fixed power assignment. The result  holds only if, for any two links having almost the same length, transmission powers  are at most a constant factor away from each other. The achieved network stability, however, is not strict Lyapunov Stability that stabilizes the system whenever the arrival rates are interior to the capacity region\cite{Neely2006ftn}. The MWISL policy we approximate can guarantee  strict Lyapunov Stability \cite{Neely2006ftn}, which indicates faster convergence and better delay performance. Meanwhile, the results on MWISL problem can further imply existence of the same order of approximation ratios for other scheduling problems (i.e., the minimum  length scheduling problem and the maximum multiflow problem)\cite{S:phy6}.

In this paper we focus on developing efficient approximation algorithms for the two variants of MWISL problem to achieve near optimal throughput region. Different from previous works that respectively develop specific algorithms for different power assignments, our work provides a unified approach to solve these variants of MWISL problem via some intrinsic properties of physical interference.
The design of our approaches is motivated by the method presented for the linear power assignment in \cite{S:phy9}. This approach works as follows. It  maps every link to a disk with a radius of double link length, and  selects a maximum weighted set of disks among the disks. Then it maps the maximum weighted set of disks back to a set of links. It proves that the interference of the link set has a constant upper bound, so the link set can be  further refined into an independent set of links. However, in contrast to their method, we applied several new designs and novel techniques.
\begin{enumerate}
  \item We utilize our characterized property of distance separation to identify candidate link sets for  refinement. Here we do not require a previously known power of any link.
  \item We design a new method to extract link sets sufficient for adjustable power assignments from these candidate link sets.
  \item We employ the fact of fading metrics \cite{halldorsson_2009} to prove those relevant results (Lemma 1 and Lemma 3). These results are independent of any power assignment.
\end{enumerate}

The main contributions  are summarized as follows.
\begin{enumerate}
  \item We characterize a kind of link set with a property of \emph{potential feasibility},  irrespective of any power assignment. We provide a sufficient condition for the property, and investigate rich features of it.
  \item Based on these fundamental results, we discover some intrinsic connections between the SINR-based and graph-based interference models regarding the MWISL problem. These connections enable us to utilize results under graph-based models to design the following two efficient scheduling algorithms for the physical interference model.
  \item For the problem with adjustable transmission power, we present a sufficient condition for feasible power assignments, and propose a $O(g(E))$-approximation algorithm for MWISL with adjustable transmission power.
  \item For the problem with fixed transmission power, we design a method applicable to any fixed transmission power case.
      It achieves $O(g(E))$ approximation when the transmission power diversity $\log\rho$ is a constant. In general it achieves $O(g(E)\log\rho)$ approximation. We further prove that the algorithm  retains  $O(g(E))$ approximation for  any  length-monotone, sub-linear fixed power assignment.
  \item We conduct extensive simulations to verify the correctness of our adjustable power assignment and evaluate throughput performance of  our proposed scheduling algorithms in various network settings. Our simulation results      demonstrate correctness and performance efficiency of our proposed algorithms. For the fixed power case when we have a previous algorithm for comparison, our proposed algorithm shows advance on throughput performance.
\end{enumerate}

The rest of the paper is organized as follows. Section \Rmnum{2} introduces the network models and problem to be studied.
Section \Rmnum{3} exploits properties on distance separation for the SINR-based interference model. Based on these properties, Section \Rmnum{4} reveals the connections between the SINR-based and graph-based interference.
Section \Rmnum{5} and \Rmnum{6}  describe our proposed algorithms for the case of adjustable and fixed transmission power respectively. Section \Rmnum{7} improves the approximation ratios and discusses distributed implementation.
\Rmnum{8} evaluates our proposed algorithms through simulations.
Section \Rmnum{9} reviews the related works. Section \Rmnum{10} concludes the paper.

\newcommand{\tabincell}[2]{\begin{tabular}{@{}#1@{}}#2\end{tabular}}
\renewcommand{\arraystretch}{1.5}
\begin{table*}[t]\setlength{\tabcolsep}{3pt}
\begin{center}
\caption{Summary of notations}
\label{notations}
\begin{tabular}{|c|c|c|c|c|c|}
  \hline
 $\eta$   & reference loss factor & $\kappa$ & path-loss exponent & $\xi$ & ambient noise\\
    \hline
 $\sigma$ & SINR threshold        & $R$      & maximum link length & $r$  & minimum link length\\
  \hline

  $ A$ &  \tabincell{c}{ the Assouad (doubling) dimension \\*(=2 for the Euclidean plane)} & $\omega$ & \tabincell{c}{the largest number of ISDs \\* that an ISL can be partitioned} &  $\epsilon$ & \tabincell{c}{a constant in $(0,1)$ \\* for the PTAS of MWISD }\\
  \hline

 $P_{\max}$ & maximum transmission power & $P_{\min}$ & minimum transmission power & $\rho$ & ratio between $P_{\max}$ and $P_{\min}$\\
 \hline
  $\delta$ & ratio between $r$ and $R$  & $C$ & a constant in doubling dimension definition & $c^{up}$ & a constant upper bound of $c_i$ \\
  \hline
\end{tabular}
\end{center}
\end{table*}

\section{Models and Preliminaries}
\subsection{Network model}
We model a wireless network by a two-tuple $(V,E)$, where $V$  denotes the set of nodes and $E$  denotes the set of links.
We assume that all nodes  are distributed in the Euclidean plane. Each node with one radio can transmit or receive at a time.
Each directed link $l_i=(s_i,t_i)\in E$  represents a communication request from a sender  $s_i$ to a receiver $t_i$. Let $d(u,v)$  denote the Euclidean distance between node $u$ and $v$, then the length of link $l_i$ is $d(s_i,t_i)$.
The  length of  link $l_i$ satisfies $ r \le d(s_i,t_i) \le R$, where $r$ and $R$ respectively denote the shortest link length and the longest link length that ensure a successful transmission. We assume that $r$ and $R$ are known for a given network, and  let $r = \delta R$, $0<\delta \leq 1$. Taking logarithm of $1/\delta$, we get \emph{length diversity} $g(E)=\log{(R/r)}$ for link set $E$  \cite{S:phy1}.
We further assume that geographical position of every node is known.

A set of links to be scheduled simultaneously must be an independent set of links (ISL) regarding to  the underlying interference models. Such a set of links is also called a \emph{feasible scheduling set}.
Under the physical interference model, it is that  each link in the set fulfills the following SINR constraint,

\[
SINR_i \buildrel \Delta \over = \frac{p(l_i) \cdot g(s_i,t_i) }{\sum\nolimits_{l_j \in \mathcal{S}} {p(l_j) \cdot
g(s_j,t_i) +\xi } }\ge \sigma _{.}
\]

Here $p(l_i)$ is the transmitting power of link $l_i$, and the power received by  $t_i$ at a distance of $d(s_i,t_i)$ is
$ p(l_i)\cdot g(s_i,t_i) \mbox{, where } g(s_i,t_i)= \min\{\eta \cdot d(s_i,t_i)^{-\kappa}, 1\} $ is the path gain from node $s_i$ to $t_i$. The constant  $\eta $ is the reference loss factor, and $\kappa $ is the path-loss exponent which satisfies $2 < \kappa < 5$ generally. $\mathcal{S}$ is the set of links which simultaneously transmit with $l_i$. The constant $\xi >0 $ denotes the ambient noise, and $\sigma $  denotes certain threshold for correct decoding of the wanted signals.

 Typically, fixed transmission power is length-monotone (i.e., $p(l_i) \geq p(l_j) \mbox{ whenever } d(s_i,t_i) \geq d(s_j,t_j)$) and sub-linear (i.e., ${p(l_i)}\slash{d(s_i,t_i) ^{\kappa}} \leq {p(l_j)}\slash{d(s_j,t_j)^{\kappa}} \mbox{ whenever } d(s_i,t_i)\geq d(s_j,t_j) $) \cite{halldosson2012infocom}.  There are three  length-monotone and sub-linear fixed power assignments widely investigated in literature \cite{halldosson2012infocom} \cite{wan2012infocom}. The first is  the uniform power assignment, where every link transmits at the same power level. The second is the linear power assignment, where $p(l_i)$ is proportional to $d(s_i,t_i)^{\kappa}$. The third is the mean power assignment, where $p(l_i)$ is proportional to $\sqrt{d(s_i,t_i)^{\kappa}}$.

For a specific power assignment $P$, we assume each link transmits at a power level $p(l_i)$, satisfying that
{\small{\[ P_{\min} \leq p(l_i) \leq P_{\max}.\]}}
Here  $P_{\max}$  and $P_{\min}$ denote the maximum transmission power and  the minimum transmission power.
 Let $\rho=P_{\max}/P_{\min}$, and $\log\rho$ denote \emph{power diversity}.
 To ensure successful transmission, $P_{\min}$  must satisfy
{\small{$P_{\min} \geq {\sigma \xi r^{\kappa}} \slash {\eta} ,$}}
even if interference from all other concurrent transmissions is zero.

\subsection{Problem definition}

The maximum throughput link scheduling problem \cite{S:MWM1}, characterizes the supportable arrival rate vectors of links for multihop wireless networks. The standard model is described as follows.
It assumes time-slotted and synchronized wireless systems with a single frequency channel. All links are assumed to have unit capacity (i.e., a link can transmit one packet with  unit length in one time slot).
At the beginning of each time slot, packets  arrive at each link independently in a stationary stochastic process  with an average arrival rate $\lambda _i$. The vector $\overrightarrow Y(T)=\{Y_i(T)\}$ denotes the number of packets arriving at each link in time slot $T$.
Every packet arrival process $Y_i(T)$ is assumed to be i.i.d over time. We also assume all packet arrival process $Y_i(T)$ have bounded second moments and they are bound by $Y_{\max}$, i.e., $Y_i(T) \leq Y_{\max}, \forall l_i \in E$.
Let a vector $\{0,1\}^{|E|}$  denote a  feasible schedule $\overrightarrow S(T)$ at each time slot $T$, where $S_i(T)=1$ if link $i$ is active in time slot $T$ and $S_i(T)=0$ otherwise. Packets departure transmitters of activated links at the end of time slots.
Then, the \emph{queue length} (it is also referred to as \emph{weight} or \emph{backlog}) vector $\overrightarrow Q(T)=\{Q_i(T)\}$ evolves according to the following dynamics:
\[
    \overrightarrow Q(T+1) = \max\{\overrightarrow 0, \overrightarrow Q(T)-\overrightarrow S(T)\}+ \overrightarrow Y(T).
\]

Described by the set of arrival rate vectors  under which the system is stable (i.e., all queues are kept finite), the\emph{ throughput capacity} (\emph{capacity region}), is a major benchmark on throughput performance.  A scheduling policy is \emph{stable}, if for any arrival rate vector in its capacity region \cite{S:pick3},
\[    \lim_{ T\rightarrow \infty} \mathbb{E}[\overrightarrow {Q}(T)] < \infty.\]

A \textit{throughput-optimal} link scheduling algorithm can achieve the \emph{optimal capacity region}, which is the union of the capacity
regions of all scheduling policies\cite{S:GMS},
\vspace{-1pt}
{\small{
\[\Lambda \triangleq \left\{ {\overrightarrow Y :\overrightarrow Y
\preccurlyeq \overrightarrow \varphi ,\mbox{ for some }\overrightarrow
\varphi \in \mbox{C}\mbox{o}\mbox{(}\Omega \mbox{)}} \right\}.
\]}}
Here $\preccurlyeq $ denotes element-wise less inequality.
$\Omega $ is the set of all feasible maximal schedules on $E$, and
$\mbox{C}\mbox{o}\mbox{(}\Omega \mbox{)}$ is the convex hull of $\Omega$.

Though we have already known that the policy of finding a MWISL at every time slot achieves the optimal capacity region, unfortunately, finding a MWISL itself is NP-hard typically \cite{sharma2006complexity}.
Thus we have to rely on approximation or heuristic methods to develop sub-optimal/imperfect scheduling algorithms running in polynomial time.

A \emph{sub-optimal scheduling policy} \cite{S:GMS} achieves a fraction of the optimal capacity region, which is characterized by \emph{efficiency ratio} $\gamma$ ($0 < \gamma \leq 1$) , i.e.,
\[\gamma \triangleq \sup \left\{ {\gamma \vert \mbox{ the network is stable
for all }\overrightarrow Y \in \gamma \Lambda } \right\}.\]

A class of imperfect scheduling policy  $\mathcal{F}_{\gamma}$ \cite{lin2005info}  is to find a set of links with a total weight that is at least $\gamma$ times of the maximum possible weight, as illustrated in the  proposition bellow.

\newtheorem{proposition}{Proposition}
\begin{proposition}(\cite{lin2005info}):
Fix $\gamma \in (0,1]$. If the user rates $\overrightarrow Y $ lie strictly inside $\gamma \Lambda$ (i.e., $\overrightarrow Y$ lies in the interior of $\gamma \Lambda $ ), then any imperfect scheduling policy  $\mathcal{F}_{\gamma}$ can stabilize the system.
\end{proposition}

Consequently, in the rest of the paper we focus on near optimal solutions to the two following variants of MWISL problem with the SINR-based physical interference constraint. Given a set $L=\{l_1, l_2,...,l_n\}$ of links,  at time slot $T$ each link $l_i$ associates with a weight $W_i(T)=Q_i(T)$, then

\begin{itemize}
  \item \textbf{MWISL problem with adjustable power}  is to find a set of links $\mathcal{S}$ with maximum total weight, and then devise a method to dynamically assign transmission power to every link in $\mathcal{S}$ such that $\mathcal{S}$ is feasible under SINR constraint;

  \item \textbf{MWISL problem with fixed power} is to find a maximum weighted independent set of links $\mathcal{S}$ regarding to the predefined transmission power of every link.
\end{itemize}

\section{Properties of distance separation}

 Given a set $L$ of links, let $V(L)$ be the set of nodes containing all senders and receivers of links in $L$. For any node $v \in V(L)$, if it holds that
{\small{
\[
    \sum \limits_{w \in V(L)} {\frac{R^{\kappa}}{d(w,v)^{\kappa}} \leq \phi}
\]}}
where $\phi$ is a constant, then we refer to such a set as a \emph{$\phi$-separation set}. For convenience, we let the element ${1} \slash {d(v,v)^{\kappa}}$ be zero.

We will show later that a $\phi$-separation set has good potential to be a feasible set.
Since any two links of a feasible scheduling set can not share a common node in a wireless network with single channel and single radio, we also suppose that  links of a $\phi$-separation set do not  share common nodes. This assumption does not influence our results.
We now introduce a sufficient condition for a $\phi$-separation set.

\newtheorem{lemma}{Lemma}

\begin{lemma}
Given a set $L$ of links,
if the distance between any two nodes of $V(L)$ is at least $d= \theta   R$,  where $\theta>0$ is a constant,
then for arbitrary  node $v \in V(L)$, we have
{\small{
\[
    \sum \limits_{w \in V(L)} {\frac{R^{\kappa}}{d(w,v)^{\kappa}} \leq \phi},
\]}}
where $\phi = \frac{ 2^{2\kappa+1} \sqrt{3} \pi   \kappa }{6 (\kappa-2) \theta^{\kappa} }$.
\end{lemma}
\begin{IEEEproof}
We leave the proof  in Appendix A for a better flow of the paper.
\end{IEEEproof}

Through the above lemma, we further conclude the following property for  a $\phi$-separation set.
\newtheorem{corollary}{Corollary}
\begin{corollary}
Given a $\phi$-separation set $L$, if it satisfies that any two nodes in $V(L)$ have a mutual distance of at least $d=\theta  R$, then for any link $l=(s_i,t_i) \in L$, it holds that
{\small{
\begin{eqnarray*}
 \sum \limits_{(s_j,t_j) \in L} \frac{d(s_i,t_i)^{\kappa}}{d(s_j,t_i)^{\kappa}} &\leq& \phi, \\
 \sum \limits_{(s_j,t_j) \in L} \frac{d(s_j,t_j)^{\kappa}}{d(s_i,t_j)^{\kappa}} & \leq& \phi, \\
 \sum \limits_{(s_j,t_j) \in L} \frac{d(s_j,t_j)^{\kappa}}{d(s_j,t_i)^{\kappa}} &\leq &\phi.
\end{eqnarray*}}}
\end{corollary}
\begin{IEEEproof}
  The results follow directly Lemma 1.
\end{IEEEproof}

\begin{lemma}
A  $\phi_1$-separation set can be partitioned into constantly many $\phi_2$-separation sets, where $\phi_1 > \phi_2$.
\end{lemma}

\begin{IEEEproof}
The proof process is similar with Theorem 1 of \cite{S:phy8}.
 \end{IEEEproof}

\section{Bridging the SINR-based and graph-based interference }

Based on those properties of distance separation, we then reveal some intrinsic  connections between the SINR-based and graph-based interference regarding  the classical MWISL problem. Through a subtle mapping between the links and interference disks, we can  obtain a set not only having the property of distance separation, but also the set having an weight of constant approximation to the optimal.

The bridging mechanism,  shown in Algorithm 1, is designed as follows.  Given a set $L$ of links, we map every link $l_i$ to a disk $a_i$. Each disk is centered at the sender of $l_i$,  with a radius of $\alpha \cdot d(s_i,t_i) \mbox{ , } \alpha > 1.$
The disk also has the same weight as that of  $l_i$. A set  of disks is independent if any pair of disks in it do not intersect with each other. We select a maximum weighted independent set of disks (MWISD) among the disks, referred to as $\mathcal{D}$.  Each disk in $\mathcal{D}$ is mapped back to the original link, and then these links compose a new set $L_{\mathcal{D}}$.

\renewcommand{\algorithmicrequire}{\textbf{Input:}}
\renewcommand{\algorithmicensure}{\textbf{Output:}}
\begin{algorithm}[t]
\caption{Bridging}
\label{alg1}
\begin{algorithmic}[1]
    \REQUIRE {Set of Links $L=\{l_1,l_2,...,l_{n}\}$}.
    \FOR {$i=1,...,n$}
        \STATE Map $l_i$ to a disk $a_i$  centered at $s_i$ with a radius of $\alpha \cdot d(s_i,t_i)$ and a weight of $W(\{l_i\})$;
     \ENDFOR
     \STATE Find a maximum weighted ISD out of the mapped disks;
     \STATE Let $\mathcal{D}$ be the selected MWISD;
     \FOR {$j=1,...,|\mathcal{D}|$}
        \STATE Map $a_j$ back to  $l_j$;
     \ENDFOR
     \STATE Let $L_{\mathcal{D}}$ be the set of links mapped back;
     \STATE Return $L_{\mathcal{D}}$.
    \end{algorithmic}
\end{algorithm}
The above procedure gracefully connects the  graph-based interference models  and SINR-based interference models. As is known there are many exiting good approximations for solving the MWISD problem (e.g., the PTAS in\cite{S:ptas})\footnote{Other constant approximation algorithms on the MWISD problem are also applicable, depending on the desired tradeoff between computation complexity and approximation ratio. For example, we can  also use other algorithms with lower complexity and smaller constant approximation ratios.}, while by the bridging mechanism we can leverage these results to solve the MWISL problem under the SINR constraint indirectly.
On one hand, by explicitly setting the radius greater than the link length, it ensures that the candidate link set satisfies the sufficient condition of a $\phi$-separation set. An appreciate $\phi$-separation set is a latent independent set of links. On the other hand, the selection of MWISD can preserve a constant-approximation ratio to the optimal under the SINR constraint, providing fundamentals  to  theoretical proofs  in later sections. The lemma below formally claims the result, irrespective of any power assignment.

\begin{lemma}
Let $L$ be an independent set of links with minimum link length of $r$ under the physical interference model, then $L$ can be partitioned into at most $\omega$ independent sets of disks with a radius of $\alpha \cdot d(s_i,t_i)$. Here $\omega=O(1/\delta^A)$.
\end{lemma}

\begin{IEEEproof}
By the technique of signal strengthening \cite{S:phy8}, we decompose the link set $L$ into $\lceil 2 \cdot 3^{\kappa} \slash \sigma \rceil ^2$ disjoint sets, each feasible with an SINR threshold of $\sigma ' = 3^{\kappa}$. We then prove that each of the sets can be partitioned into constantly many independent set of disks with a radius of $\alpha \cdot d(s_i, t_i)$. For convenience, we let $3^{\kappa} > \sigma$,  but we do not necessarily assume $3^{\kappa} > \sigma$, we later show that the lemma also holds when $3^{\kappa} \leq \sigma$.

\newtheorem{claim}{Claim}
\begin{claim} \cite{Kesselheim2011soda}
 Considering any two distinct links $l_i=(s_i,t_i) \mbox{ , } l_j=(s_j,t_j)$ in one of the sets $L'$ fulfilling the SINR threshold $\sigma ' = 3^{\kappa}$, the distance between any  pair of the involved nodes $s_i \mbox{ , } t_i \mbox{ , } s_j \mbox{ , } t_j$ has to be at least $r$ for any power setting.
\end{claim}

It is obvious that the claim still holds when $3^{\kappa} \leq \sigma$, and nodes in $V(L')$ have a mutual distance of $r$ at least.

We then prove that the corresponding disk of any link  in $L'$ intersects constantly many disks of other links. We observe that any pair of disks intersect if and only if the mutual distance of their senders are less than $\alpha \cdot d(s_i,t_i) + \alpha \cdot d(s_j,t_j)$. We then get that no disks will intersect with other disks if the distance of any two distinct senders is above $2 \alpha R$.
Therefore, we just need to show only a constant number of senders located in the disk centered at any sender  with a radius of $2 \alpha R$.

We initially assume that set $V_s(L')$ consists of senders of all links in $L'$.
For any sender $s_i \in V_s(L')$, we define a following set fulfilling,
{\small{
\[V' = \biggl\{s_j \in V_s(L') | d(s_j,s_i) \leq \frac{2 \alpha R}{r \slash 2} \cdot \frac{r}{ 2}\biggl\}.\]
}}
It is  obvious that $|V'|$ is the maximum possible  number of links whose disks may intersect with the disk of the given link $l_i$.
To prove it, we just need to apply the fact of fading metrics and the packing bound once again.

By Claim 1, we have already known that the distance between any pair of distinct senders is at least $r$. This is, balls of radius $r \slash 2$ centered at nodes in $V'$  are fully contained in $B \left(s_i,  (\frac{2 \alpha R}{r \slash 2}+1) \cdot \frac{r}{ 2}\right)$. It implies
{\small{
\[|V'| \leq C \biggl(\frac{2 \alpha R}{r \slash 2}+1 \biggl)^A=  C \biggl(\frac{4 \alpha}{\delta}+1 \biggl)^A,\]
}}
and we get the number of disks which may  intersect with the same disk bounded.

Thus $L$ can be at most partitioned into
\[\omega= \lceil 2 \cdot 3^{\kappa} \slash \sigma \rceil ^2 \cdot |V'|=O(1/\delta^A)=O(1/\delta^2)\]
subsets of links, the disks of which do not intersect mutually.

This finishes the proof.
\end{IEEEproof}

Note that our proposed algorithm and results apply equally to the bi-directional link case. In a bi-directional case, each node is a sender and receiver. That is, for each bi-directional link $l_i$ with endpoints $u$ and $v$, $u$ gets $w_u$ packets to $v$ while $v$ gets $w_v$ packets to $u$. We just need to change the previous one-to-one mapping in Algorithm 1 to a one-to-two mapping. We take link $l_i$ as two directional links, $(u,v)$ with a weight of $w_u$ and $(v,u)$ with a weight of $w_v$. According to the mapping rules in Algorithm 1,  we get two distinct disks with the same radius of $2\|l_i\|$, one centered at $u$ with a weight of $w_u$, the other centered at $v$ with a weight of $w_v$. This process just doubles the number of candidate disks for computing of MWISD, but does not influence the derivation of a constant-approximation or logarithmic-approximation ratio for the optimal.

\section{Approximation algorithms with adjustable transmission power}

Now we utilize these fundamentals to develop solutions to the MWISL with adjustable transmission power.

\subsection{Adjustable power assignment}
Given a feasible scheduling set $L$, to ensure feasibility, the assigned power $p(l_i)$ for link $l_i$ shall satisfy,
{\small{
\[
    p(l_i)\geq
              \sigma  \cdot \left( \frac{\xi}{\eta} {d(s_i,t_i)^{\kappa}}  +
                   \sum \limits_{l_j \in \{L \backslash l_i\}} p(l_j)\frac{d(s_i,t_i)^{\kappa}}{d(s_j,t_i)^{\kappa}}  \right).
\]
}}
That is, the assigned power shall compensate the interference of ambient noise and the simultaneous transmissions.
Therefore, a natural method is to assign the power iteratively, and  compensate the interference respectively from ambient noise, the previously assigned links and the later assigned links. For a specific link, the suffered interference from ambient noise and the previously assigned links is easy to calculate, the question is to estimate an upper bound of the interference from the later assigned links. Works in \cite{Kesselheim2011soda},\cite{wan2011maximizing} provide hints for the problem. The basic idea is to assign a power that is a scale of the summed interference from ambient noise and the preassigned links. Then the interference from the later assigned links can  be actually looked upon as an indirect interference of the noise and previously assigned links. However, the sufficient conditions on the candidate link set for feasible power assignments differ greatly. We first introduce the procedure of power assignment below.

\textbf{Iterative power assignment.} Consider a set $L^*$ of links, and let $l_1,l_2,...,l_n$ be a permutation of links in $L^*$.
Note in this procedure, we do not need to order the links in the ordering of decreasing length as done in  \cite{Kesselheim2011soda},\cite{wan2011maximizing}. It works for arbitrary ordering of links as long as the given set satisfies the sufficient condition we propose later.
Assign the first link $l_1$ a power,
{\small{\[
  p(l_1) = m  \frac{\sigma \xi}{\eta} d(s_1,t_1)^{\kappa}.
\]}}
The powers assigned to later links are iteratively set  by
{\small{
\[
  p(l_i) = m \sigma d(s_i,t_i)^{\kappa} \left( \sum^{i-1} \limits_{j=1} \frac{p(l_j)}{d(s_j,t_i)^{\kappa}} + \frac{\xi}{\eta} \right).
\]
}}
Here $m$ relates to the distance separation property of the given set.
Certainly, $m$ shall be greater than $1$ to cover the interference from the ambient noise and the previously assigned links. A strict bound of $m$ depends on the sufficient conditions below.

\begin{lemma}
    Let $L^*$ be a $\phi$-separation set, if $L^*$ fulfills the two following conditions simultaneously:
     \begin{enumerate}
      \item for any two distinct links in $L^*$, i.e., $l_i$, $l_j$, $i \neq j$, the mutual distance between the senders $s_i$, $s_j$ is at least $\alpha \cdot (d(s_i,t_i)+d(s_j,t_j))$,
      \item the constant $\phi$ satisfies that $\phi \leq \frac{1}{4\cdot {\beta}^{\kappa} \sigma (\sigma+1)}$, where $\beta=\frac{2\alpha-1}{\alpha-1}$.
     \end{enumerate}
    then the iterative power assignment generates a feasible power assignment for $L^*$ if $m$ is within
{\small{
       \[  \left[\frac{1-\sqrt{1-4\cdot {\beta}^{\kappa} \phi \sigma (\sigma+1)}}{2 \cdot {\beta}^{\kappa} \phi  \sigma (\sigma+1)},\frac{1+\sqrt{1-4\cdot {\beta}^{\kappa} \phi \sigma (\sigma+1)}}{2 \cdot {\beta}^{\kappa} \phi  \sigma (\sigma+1)}\right].\]
}}
\end{lemma}

\begin{IEEEproof}
The proof respectively treats the interference from the ambient noise, the previously assigned links and the later assigned links.
It is equivalent to prove that
{\small{
\begin{eqnarray*}
  p(l_i) &= & m \sigma d(s_i,t_i)^{\kappa} \left( \sum^{i-1} \limits_{j=1} \frac{p(l_j)}{d(s_j,t_i)^{\kappa}} + \frac{\xi}{\eta} \right) \\*
    &\geq&
              \sigma  \cdot \biggl( \frac{\xi}{\eta} {d(s_i,t_i)^{\kappa}}  +
                   \sum^{i-1}\limits_{j=1} p(l_j)\frac{d(s_i,t_i)^{\kappa}}{d(s_j,t_i)^{\kappa}}      \\
       &~&{}   +           \sum^{n} \limits_{j=i+1} p(l_j)\frac{d(s_i,t_i)^{\kappa}}{d(s_j,t_i)^{\kappa}} \biggl).
\end{eqnarray*}
}}
By rearranging the terms, we just need to show that the interference from the later assigned links get bounded by,
{\small{
\begin{eqnarray*}
    \sum^{n} \limits_{j=i+1} \frac{p(l_j)}{d(s_j,t_i)^{\kappa}}
  &\leq&
       (m-1) \left( \sum^{i-1} \limits_{j=1} \frac{p(l_j)}{d(s_j,t_i)^{\kappa}} + \frac{\xi}{\eta} \right) \\
  &=&  \frac{m-1}{m\sigma} \cdot \frac{p(l_i)}{d(s_i,t_i)^{\kappa}}.
\end{eqnarray*}
}}
As we mentioned previously, the basic idea of the proof is to take the interference from the later assigned links as the indirect effect of the interference from the previously added. A crucial claim to bound the indirect interference is as follows.

For any $i$ and $k$ within $\max\{i,k\} < n$,
{\small{
\[
    \sum^{n} \limits_{j=\max\{i,k\} } \frac{d(s_j,t_j)^{\kappa} \cdot d(s_k,t_i)^{\kappa}}
                                         {d(s_k,t_j)^{\kappa} \cdot d(s_j,t_i)^{\kappa}} \leq \left(\frac{2\alpha-1}{\alpha-1} \right)^{\kappa} \cdot \phi.
\]
}}
From the first condition of $L^*$, we can get that $d(s_k,t_j)> (\alpha-1) \cdot d(s_j,t_j)$, since
{
    \small{\begin{eqnarray*}
  d(s_k,t_j) &\geq& d(s_k,s_j)-d(s_j,t_j)  \\*
             &\geq& \alpha \cdot (d(s_k,t_k)+d(s_j,t_j)) -d(s_j,t_j) \\*
             &=& \alpha \cdot d(s_k,t_k)+ (\alpha-1) \cdot d(s_j,t_j).
\end{eqnarray*}
    }
}
Similarly, we get that $d(s_j,t_i)> (\alpha-1) d(s_j,t_j)$ by
{\small{
\begin{eqnarray*}
  d(s_j,t_i) &\geq& d(s_j,s_i)-d(s_i,t_i)  \\*
             &\geq& \alpha \cdot (d(s_i,t_i)+d(s_j,t_j)) -d(s_i,t_i) \\*
             &=& \alpha \cdot d(s_j,t_j)+ (\alpha-1) d(s_i,t_i )\\*
             & >& (\alpha-1) d(s_j,t_j).
\end{eqnarray*}
}}
Having the two inequalities, it then follows,
{\small{
\begin{eqnarray*}
   &~& \sum^{n} \limits_{j=\max\{i,k\} } \frac{d(s_j,t_j)^{\kappa} \cdot d(s_k,t_i)^{\kappa}}
                                         {d(s_k,t_j)^{\kappa} \cdot d(s_j,t_i)^{\kappa}}       \\*
   &\leq& \sum^{n} \limits_{j=\max\{i,k\} } \frac{d(s_j,t_j)^{\kappa} \cdot
                                         (d(s_k,t_j)+d(s_j,t_j)+d(s_j,t_i))^{\kappa}}
                                         {d(s_k,t_j)^{\kappa} \cdot d(s_j,t_i)^{\kappa}}       \\*
   &\leq& \sum^{n} \limits_{j=\max\{i,k\} } d(s_j,t_j)^{\kappa} \cdot \left(\frac{d(s_k,t_j) + \frac{1}{2} d(s_j,t_j)} {d(s_k,t_j) \cdot d(s_j,t_i)}           +                          \right.  \\*
   & &  \left.  \frac{d(s_j,t_i)+\frac{1}{2}d(s_j,t_j)}{d(s_k,t_j) \cdot d(s_j,t_i)}  \right)^{\kappa} \\
       \end{eqnarray*}}}
 {\small{
 \begin{eqnarray*}
   &\leq& \sum^{n} \limits_{j=\max\{i,k\} } d(s_j,t_j)^{\kappa} \cdot \left(\frac{d(s_k,t_j) + \frac{1}{2 (\alpha-1)}  d(s_k,t_j)} {d(s_k,t_j) \cdot d(s_j,t_i)}           +                          \right.  \\*
   & &  \left.  \frac{d(s_j,t_i)+\frac{1}{2 (\alpha-1)}d(s_j,t_i)}{d(s_k,t_j) \cdot d(s_j,t_i)}  \right)^{\kappa} \\
   &\leq& \sum^{n} \limits_{j=\max\{i,k\} } \left(\frac{2\alpha -1}{2(\alpha-1)}\right)^{\kappa} \cdot \left( \frac{d(s_j,t_j)}{d(s_j,t_i)} + \frac{d(s_j,t_j)}{d(s_k,t_j)}\right)^{\kappa}                                                  \\
   &\leq& \sum^{n} \limits_{j=\max\{i,k\} } \left(\frac{2\alpha -1}{2(\alpha-1)}\right)^{\kappa} 2^{\kappa -1} \cdot \left[
    \left( \frac{d(s_j,t_j)}{d(s_j,t_i)} \right)^{\kappa}+ \left(\frac{d(s_j,t_j)}{d(s_k,t_j)} \right)^{\kappa}\right] \\
   &\leq& \frac{(2\alpha -1)^{\kappa}}{2(\alpha-1)^{\kappa}} \left[ \sum^{n} \limits_{j=\max\{i,k\}} \left( \frac{d(s_j,t_j)}{d(s_j,t_i)} \right)^{\kappa} +\sum^{n} \limits_{j=\max\{i,k\}} \left(\frac{d(s_j,t_j)}{d(s_k,t_j)} \right)^{\kappa} \right]             \\
   &\leq& \left(\frac{2\alpha-1}{\alpha-1} \right)^{\kappa} \phi \mbox{ .}
\end{eqnarray*}
}}
The last third inequality results from the generalized mean inequality (cf. \cite{wan2011maximizing}). The last step is built on Corollary 1. For brevity, we let $\frac{2\alpha-1}{\alpha-1} =\beta$.

Next we analyze the upper bound of the interference from the later assigned links.
{\small{
\begin{eqnarray*}
  &~&    \sum^{n} \limits_{j=i+1}\frac{ p(l_j)}{d(s_j,t_i)^{\kappa}} \\
  &=&    \sum^{n} \limits_{j=i+1}  m \sigma \left( \sum^{j-1} \limits_{k=1} \frac{p(l_k) \cdot d(s_j, t_j)^{\kappa} }
                                                                         {d(s_k, t_j)^{\kappa} \cdot d(s_j,t_i)^{\kappa} }
           + \frac{ \xi}{\eta} \frac{d(s_j,t_j)^{\kappa}} {d(s_j,t_i)^{\kappa}} \right) \\
  &=&  m \sigma\left(  \sum^{n} \limits_{j=i+1}   \sum^{j-1} \limits_{k=1} \frac{p(l_k) \cdot d(s_j, t_j)^{\kappa} }
                                                                         {d(s_k, t_j)^{\kappa} \cdot d(s_j,t_i)^{\kappa} }
           + \frac{ \xi}{\eta} \sum^{n} \limits_{j=i+1}  \frac{d(s_j,t_j)^{\kappa}} {d(s_j,t_i)^{\kappa}}  \right)\\
  &\leq&  m \sigma \sum^{n} \limits_{j=i+1} \sum^{j-1} \limits_{k=1}  \frac{p(l_k) \cdot d(s_j, t_j)^{\kappa} }
                                                                         {d(s_k, t_j)^{\kappa} \cdot d(s_j,t_i)^{\kappa} }
         + m \sigma \phi \frac{\xi}{\eta} \mbox{ .}
\end{eqnarray*}
}}
 The last inequality bases on the third inequality in Corollary 1. We then focus on the first term of the above inequality, by rearranging the sums we get
 {\small{
 \begin{eqnarray*}
 &~& m \sigma \sum^{n} \limits_{j=i+1} \sum^{j-1} \limits_{k=1}  \frac{p(l_k) \cdot d(s_j, t_j)^{\kappa} }
                                                                  {d(s_k, t_j)^{\kappa} \cdot d(s_j,t_i)^{\kappa} } \\
 &\leq&  m \sigma \sum^{n} \limits_{j=i+1} \sum^{i} \limits_{k=1}  \frac{p(l_k) \cdot d(s_j, t_j)^{\kappa} }
                                                                  {d(s_k, t_j)^{\kappa} \cdot d(s_j,t_i)^{\kappa} } + \\
 &~& m \sigma \sum^{n} \limits_{j=i+1} \sum^{n} \limits_{k=i+1}  \frac{p(l_k) \cdot d(s_j, t_j)^{\kappa} }
                                                                  {d(s_k, t_j)^{\kappa} \cdot d(s_j,t_i)^{\kappa} } \\
 &=&  m \sigma \sum^{i} \limits_{k=1} \sum^{n} \limits_{j=i+1}  \frac{p(l_k) \cdot d(s_j, t_j)^{\kappa} }
                                                                  {d(s_k, t_j)^{\kappa} \cdot d(s_j,t_i)^{\kappa} } + \\
 &~&  m \sigma \sum^{n} \limits_{k=i+1} \sum^{n} \limits_{j=i+1}  \frac{p(l_k) \cdot d(s_j, t_j)^{\kappa} }
                                                              {d(s_k, t_j)^{\kappa} \cdot d(s_j,t_i)^{\kappa} } \\
 &=&  m \sigma \sum^{i} \limits_{k=1} \frac{p(l_k) }{d(s_k,t_i)^{\kappa}}
                                \sum^{n} \limits_{j=i+1}  \frac{d(s_k,t_i)^{\kappa} \cdot d(s_j, t_j)^{\kappa} }
                                                                  {d(s_k, t_j)^{\kappa} \cdot d(s_j,t_i)^{\kappa} }  + \\*
 &~& {} m \sigma \sum^{n} \limits_{k=i+1} \frac{p(l_k) }{d(s_k,t_i)^{\kappa}}
                             \sum^{n} \limits_{j=i+1}  \frac{d(s_k,t_i)^{\kappa} \cdot d(s_j, t_j)^{\kappa} }
                             {d(s_k, t_j)^{\kappa} \cdot d(s_j,t_i)^{\kappa} } \\
  &\leq&  m \sigma \sum^{i} \limits_{k=1} \frac{p(l_k) \cdot {\beta}^{\kappa}  \phi}{d(s_k,t_i)^{\kappa}}
      + m \sigma \sum^{n} \limits_{k=i+1} \frac{p(l_k)  \cdot {\beta}^{\kappa}  \phi}{d(s_k,t_i)^{\kappa}} \\
  \end{eqnarray*}}}
 {\small{
 \begin{eqnarray*}
 &\leq&  m \sigma {\beta}^{\kappa} \phi \sum^{i-1} \limits_{k=1} \frac{p(l_k) }{d(s_k,t_i)^{\kappa}}  +
                  m \sigma  {\beta}^{\kappa}  \phi   \frac{p(l_i) }{d(s_i,t_i)^{\kappa}}         +         \\
 &~& {}m \sigma {\beta}^{\kappa}  \phi  \sum^{n} \limits_{k=i+1} \frac{p(l_k) }{d(s_k,t_i)^{\kappa}} \\
 &=& (1+m \sigma) {\beta}^{\kappa}  \phi  \frac{p(l_i) }{d(s_i,t_i)^{\kappa}} -  m \sigma {\beta}^{\kappa}\phi \frac{\xi}{\eta}+ \\*
 &&   m \sigma {\beta}^{\kappa}\phi \sum^{n} \limits_{k=i+1} \frac{p(l_k) }{d(s_k,t_i)^{\kappa}} .
\end{eqnarray*}
}}

Thus we can surely get a bounded interference by,
{\small{
\begin{eqnarray*}
 &~&   \sum^{n} \limits_{j=i+1} \frac{p(l_j)}{d(s_j,t_i)^{\kappa}} \\
&\leq& (1+m \sigma) {\beta}^{\kappa}  \phi  \frac{p(l_i) }{d(s_i,t_i)^{\kappa}} -  m \sigma {\beta}^{\kappa}\phi \frac{\xi}{\eta} + \\
 &&  m \sigma {\beta}^{\kappa}\phi \sum^{n} \limits_{k=i+1} \frac{p(l_k) }{d(s_k,t_i)^{\kappa}} + m \sigma \phi \frac{\xi}{\eta}\\
   &\leq&  (1+m \sigma) {\beta}^{\kappa} \phi  \frac{p(l_i) }{d(s_i,t_i)^{\kappa}}
   + m \sigma {\beta}^{\kappa}\phi \sum^{n} \limits_{k=i+1} \frac{p(l_k) }{d(s_k,t_i)^{\kappa}} \\
    &\leq& \frac{(1+m \sigma) {\beta}^{\kappa} \phi }{1-m \sigma {\beta}^{\kappa}\phi} \cdot \frac{p(l_i) }{d(s_i,t_i)^{\kappa}} ,
\end{eqnarray*}
}}
because $m \sigma {\beta}^{\kappa}\phi \leq \frac{1}{1+\sigma}<1$ when  $m$ in
{\small{
\[  \left[\frac{1-\sqrt{1-4\cdot {\beta}^{\kappa} \phi \sigma (\sigma+1)}}{2 \cdot {\beta}^{\kappa} \phi  \sigma (\sigma+1)},\frac{1+\sqrt{1-4\cdot {\beta}^{\kappa} \phi \sigma (\sigma+1)}}{2 \cdot {\beta}^{\kappa} \phi  \sigma (\sigma+1)}\right].\]
}}
Finally we can confirm that when $m$ lies in the region, combining $\phi \leq \frac{1}{4\cdot {\beta}^{\kappa}  \sigma (\sigma+1)}$,  it exactly ensures the following inequality holds, which is also the final objective of this proof,
{\small{
\begin{eqnarray*}
  \sum^{n} \limits_{j=i+1} \frac{p(l_j)}{d(s_j,t_i)^{\kappa}} &\leq& \frac{(1+m \sigma) {\beta}^{\kappa} \phi }{1-m \sigma {\beta}^{\kappa}\phi} \cdot \frac{p(l_i) }{d(s_i,t_i)^{\kappa}} \\
&\leq& \frac{m-1}{m\sigma} \cdot \frac{p(l_i)}{d(s_i,t_i)^{\kappa}} .
\end{eqnarray*}
}}
This finishes the proof.
\end{IEEEproof}

\begin{lemma}
Given a $\phi$-separation set fulfilling the sufficient conditions, using  the iterative power assignment, the  assigned power  has an upper bound of
{\small{ \[P_{\max}^{up}=\frac{m \sigma \xi R^{\kappa}}{(1-m \sigma \phi) \eta}.\]}}
\end{lemma}
\begin{IEEEproof}
From Lemma 4, we have $m \sigma \phi {\beta}^{\kappa} <1$.
We then prove by induction.
For the first assigned link, it holds that
{\small{\[
  p(l_1)  =  \frac{m\sigma \xi}{\eta} d(s_1,t_1)^{\kappa}
        \leq \frac{m\sigma \xi}{\eta} R^{\kappa}
        = (1-m\sigma \phi)P_{\max}^{up}.
\]}}

If for any later assigned link $l_i, i>1$, $p(l_i) \leq P_{\max}$, then for $l_{i+1}$ we still have
{\small{
 \begin{eqnarray*}
  p(l_{i+1})& =& m \sigma d(s_{i+1},t_{i+1})^{\kappa} \left( \sum^{i} \limits_{j=1} \frac{p(l_j)}{d(s_j,t_{i+1})^{\kappa}} + \frac{\xi}{\eta} \right)                                                                        \\
  &\leq&  m \sigma \phi P_{\max}^{up} + {(1-m \sigma \phi)} P_{\max}^{up}
 {~} = {~} P_{\max}^{up}.
\end{eqnarray*}
}}
This finishes the proof.
\end{IEEEproof}

Remember that $P_{\max}$ refers to the maximum transmission power of all links, thus it also satisfies that $P_{\max} \leq P_{\max}^{up} $.

\subsection{Approximation algorithm}
 Now we describe our proposed algorithm for MWISL with adjustable transmission power.  The pseudo codes are shown in Algorithm 2.

\begin{algorithm}[t]
\caption{Approximation algorithm with adjustable power}
\label{alg2}
\begin{algorithmic}[1]
    \REQUIRE {Set of Links $E=\{l_1,l_2,...,l_{|E|}\}$}.
    \STATE Preprocess $E$ using the Bridging algorithm and let  $L_{\mathcal{D}}$ be the returned result;
     \STATE Refine $L_{\mathcal{D}}$ to a collection of $\phi^*$-separation sets with $\phi^* = \frac{1}{4\cdot {\beta}^{\kappa}  \sigma (\sigma+1)}$ by the first-fit algorithm in proof of Lemma 2;
     \STATE Select the most weighted set from the collection;
     \STATE Let $L_{\mathcal{D}}^*=\{l_1^*,l_2^*, ... ,l_n^*\}$ be the resulted set;
     \STATE Assign power to each link of $L_{\mathcal{D}}^*$ by the iterative power assignment with $m=2$;
     \STATE Return $L_{\mathcal{D}}^*$ with assigned powers.
    \end{algorithmic}
\end{algorithm}

\newtheorem{theorem}{Theorem}
\begin{theorem}
Algorithm 2 for MWISL with adjustable transmission power outputs a feasible scheduling set having a weight of $O(1/\delta^{2(\kappa+1)})$ approximating to the optimal.
\end{theorem}

\begin{IEEEproof}
We first verify the correctness of the algorithm.

It is obvious that $L_{\mathcal{D}}^*$ fulfills the sufficient conditions for a feasible power assignment.  And $\phi^*= \frac{1}{4\cdot {\beta}^{\kappa}  \sigma (\sigma+1)}$ makes $m=2$ exactly. Thus, the iterative power assignment generates  a feasible power assignment for  $L_{\mathcal{D}}^*$.

Then we prove the theoretical bound for the algorithm.
We use $W(L)$ to denote the summed weight of a set $L$, and $W(OPT)$ to denote  the optimum.

The nodes of all links in $L_{\mathcal{D}}$ have a smallest mutual distance of $r=\delta  R$. Thus, according to Lemma 1, $L_{\mathcal{D}}$  is a
$\phi'$-separation set, where
{\small{\[   \phi' =  \frac{ 2^{2\kappa+1} \sqrt{3} \pi   \kappa }{6 (\kappa-2) \delta^{\kappa} } .\]}}

Next, by Lemma 2, $L_{\mathcal{D}}$ can be partitioned into at most $\omega_1$ $\phi^*$-separation sets, where $\omega_1$ is a constant upper bounded by
{\small{
\begin{eqnarray*}
    \omega_1 &= & 4 \cdot \lceil \frac{\phi'^2}{{\phi^*}^2}\rceil \\
    &\leq& 4 \cdot \biggl[  \frac{ 2^{2\kappa+1} \sqrt{3} \pi   \kappa }{6 (\kappa-2) \delta^{\kappa} } \cdot {4\cdot {\beta}^{\kappa}  \sigma (\sigma+1)}\biggl]^2 \\
    &=&   4^{2\kappa+3} {\pi}^2 {\beta}^{2\kappa} \left[ \frac{    \kappa \sigma (\sigma+1)} {\delta^{\kappa}(\kappa-2)}\right]^2  / 3.
\end{eqnarray*}
}}

For $L_{\mathcal{D}}^*$ is the most weighted one among the collection, we further have
{\small{\[\omega_1  \cdot  W(L_{\mathcal{D}}^*) \geq W(L_{\mathcal{D}}).\]}}
By Lemma 3, any feasible set of links can be partitioned into at most $\omega$ ISDs, so the optimal MWISL has a weight at most
 $\omega \cdot W(L_{\mathcal{D}}) \slash (1-\epsilon)$. Here $1 \slash (1-\epsilon)$ is the approximation ratio of Algorithm 2 in \cite{S:ptas}  for the MWISD problem.

Consequently, we get
{\small{\[
   \frac{\omega  \omega_1}{1-\epsilon}   W(L_{\mathcal{D}}^*) \geq W(OPT).
\]}}
This completes the proof.
\end{IEEEproof}

 We then analyze the time complexity of Algorithm 2. The algorithm mainly consists of the bridging process, the refinement process and power assignment part. The refinement process and the iterative power assignments respectively cost  $O(|E|)$ rounds. The complexity of the bridging process depends on the graph-based algorithm for MWISD problem. If we utilize PTAS\cite{S:ptas} in this part, the complexity would be exponential of $|E|$. It is ok to small-scale networks, but not applicable to large-scale networks. To improve efficiency, we can choose other simple constant-approximation algorithms with some sacrifice of approximation ratios. For instance, we can use greedy maximal schedule to find MWISD in complexity of $O(|E|\log(|E|)$. Then the complexity of Algorithm 2 will be reduced to $O(|E|\log(|E|))$.

\section{Approximation algorithm with fixed transmission power}
In this section we study the problem with fixed transmission power. Similarly with Algorithm 2, Algorithm 3  is still built on our proposed properties and bridge.
We first list several existing results which facilitate a simple proof of our proposed algorithm.
\newtheorem{definition}{Definition}
\begin{definition}
(affectance \cite{S:phy8}) The relative interference of link $l_j $ on $l_i $ is the increase caused by  $l_j $ in the inverse of the SINR at  $l_i $, namely
{\small{\[r_{l_j } (l_i )
                 =\frac{p(l_j) \cdot  g(s_j,t_i)} { p(l_i) \cdot g(s_i,t_i)}.\]
}}
For convenience, define  $r_{l_i} (l_i )=0$. Let
{\small{ \[c_i =\frac{\sigma }{1-\sigma \xi \slash \left( p(l_i) \cdot g(s_i,t_i) \right) }\]
}}
indicate the extent to which the ambient noise approaches the required signal at receiver  $t_i $. Since $c_i$ is a constant related to the properties of link $l_i$, we assume a constant upper bound of $c_i$ for all links, i.e.,
 \[c^{up}= \max_{l_i \in E}\{c_i\} \leq h \sigma , h>1 .\]
This  is a fairly reasonable assumption. It simply says that in the absence of other  concurrent transmissions, the
transmission succeeds comfortably.
The affectance of link $l_i $,  caused by a set $\mathcal{S}$  of links that transmit simultaneously with $l_i$, is the sum of relative interference of the links in  $\mathcal{S}$  on  $l_i$, scaled by $c_i $, or
{\small{\[a_{\mathcal{S}} (l_i )=c_i \cdot \sum\limits_{l_j \in \mathcal{S}} {r_{l_j } (l_i )}.\]}}
For a single link $l_j$, we use the shorthand $a_j(l_i)=a_{l_j}(l_i)$.
\end{definition}

\begin{definition}
($\tau \mbox{-signal}$ set \cite{S:phy8}) We define a $\tau \mbox{-signal}$  set to be one where the affectance of any link is at most  $1 \slash \tau$. Clearly, any ISL is a $\mbox{1-signal}$ set.
\end{definition}

\begin{lemma}
$L_{\mathcal{D}}^*$ is a $\tau$-signal set, and ${1}\slash{\tau}$ is  bounded above  by ${c^{up}  \rho \phi}$ when $\rho$ is a constant and ${2 c^{up} \phi}$  otherwise.
\end{lemma}
\begin{IEEEproof}
The affectance of each link $l_i \in L_{\mathcal{D}}^*$ satisfies,
{\small{
\begin{eqnarray*}
  a_{L_{\mathcal{D}}^*}(l_i) &\leq& a_{V({L_\mathcal{D}})}(t_i) \\*
  & \leq & c_i \cdot \sum_{ \substack{w \in V({L_\mathcal{D}}) }} \left( \frac{p(l_w)}{p(l_i)} \cdot \frac{d(s_i,t_i)^{\kappa} }{d(w,t_i)^{\kappa}} \right)\\
  & \leq &  c^{up} \cdot \sum_{ \substack{w \in V({L_\mathcal{D}}) }} \left(\frac{p(l_w)}{p(l_i)} \cdot \frac{R^{\kappa} }{d(w,t_i)^{\kappa}} \right).
\end{eqnarray*}
}}

If $\rho$ is a constant then,
{\small{\[a_{L_{\mathcal{D}}^*}(l_i) \leq c^{up}  \rho \phi,\]}}
otherwise,
{\small{\[a_{L_{\mathcal{D}}^*}(l_i) \leq 2 c^{up}   \phi,\]}}
where $\phi =  \frac{ 2^{2\kappa+1} \sqrt{3} \pi   \kappa }{6 (\kappa-2) \delta^{\kappa} }$ by Lemma 1.

Therefore, we have ${1}\slash{\tau}$ bounded by ${c^{up}  \rho \phi}$ when $\rho$ is a constant and ${2 c^{up} \phi}$  otherwise.
\end{IEEEproof}

Next we give the approximation ratio for our algorithm.

\begin{algorithm}[t]
\caption{Approximation algorithm with fixed  power}
\label{alg3}
\begin{algorithmic}[1]
    \REQUIRE {Set of Links $E=\{l_1,l_2,...,l_{|E|}\}$}.
     \STATE Preprocess $E$ using the Bridging mechanism and let $L_{\mathcal{D}}$ be the output;
     \IF {$\rho$  is not a constant}
     \STATE  Divide $L_{\mathcal{D}}$ into $\log{\rho}$ sets and choose the most weighted set as $L_{\mathcal{D}}^*$;
     \ELSE
        \STATE let $L_{\mathcal{D}}^*=L_{\mathcal{D}}$;
     \ENDIF
     \STATE Refine $L_{\mathcal{D}}^*$ to a collection of ISLs using a simple first-fit greedy method;
     \STATE Select an ISL with the largest weight as $\mathcal{S}$;
     \RETURN $\mathcal{S}$.
    \end{algorithmic}
\end{algorithm}

\begin{theorem}
Algorithm 3 achieves an approximation ratio of $O(1/\delta^{2(\kappa+1)})$ for the MWISL problem with fixed transmission power when $\rho$ is a constant, and an approximation ratio of $O(\log\rho/\delta^{2(\kappa+1)})$ generally.
\end{theorem}
\begin{IEEEproof}
By the technique of signal strengthening \cite{S:phy8}, $L_{\mathcal{D}}^*$ can be partitioned into $4 \slash \tau ^2$ ISLs at most,
thus
{\small{\[\frac{4} {\tau ^2} \cdot W(\mathcal{S}) \geq  W(L_{\mathcal{D}}^*).\]
}}
By Algorithm 3, we have
{\small{\[    W(L_{\mathcal{D}}^*)       = W(L_{\mathcal{D}})\]
}}
if $\rho$ is a constant, or
{\small{ \[ \log{\rho} \cdot W(L_{\mathcal{D}}^*) \geq W(L_{\mathcal{D}})\]
}}
since the most weighted set is selected as $L_{\mathcal{D}}^*$.

Through Lemma 3, the optimal MWISL has a weight at most
 $\omega \cdot W(L_{\mathcal{D}}) \slash (1-\epsilon)$.

Hence, when $\rho $ is a constant we have,
{\small{\[\frac{4  \omega   }{ (1-\epsilon)\tau^2} \cdot W(\mathcal{S}) \geq W(OPT),\mbox{ where } \frac{1}{\tau}=c^{up} \rho \phi,\]}}
and when $\rho $ is not a constant we have,
 {\small{\[\frac{4  \omega    \log{\rho} }{(1-\epsilon)\tau^2 }  \cdot W(\mathcal{S}) \geq W(OPT), \mbox{ where } \frac{1}{\tau}=2 c^{up} \phi .\]}}
\vspace{-2px}
\end{IEEEproof}

\begin{theorem}
For any sub-linear and length-monotone fixed power assignment, e.g., the uniform power assignment, the linear power assignment, and the mean power assignment, Algorithm 3 has an approximation factor of $O(1/\delta^{2(\kappa+1)})$.
\end{theorem}
\begin{IEEEproof}
Considering any two distinct links, $l_i$ and $l_j$, we assume $d(s_i,t_i) > d(s_j,t_j)$ for brevity, then we have
{\small{
\[    {p(l_i)} \slash {p(l_j)} < {d(s_i,t_i)^{\kappa}} \slash {{d(s_j,t_j)^{\kappa}}}\]}}
 by the sub-linear feature. Thus we further get $\rho$ bounded by,
{\small{ \[ {\rho} = {P_{\max}} \slash {P_{\min}} < {R^{\kappa}} \slash {r^{\kappa}}.\]}}
Immediately, we also get
 $ {1}\slash{\tau} =  {c^{up}   \phi }\slash{ \delta^{\kappa} }$
 for the corresponding approximation ratio $\frac{4  \cdot \omega   }{(1-\epsilon)\tau^2}$.
\end{IEEEproof}

The complexity of Algorithm 3 is the same as Algorithm 2.

\section{Improving the algorithms}

\subsection{Improving approximation ratios}

For both Algorithm 2  and Algorithm 3, the approximation ratios  polynomial in $1/\delta=R/r$  could be further improved to logarithmic of $R/r$ by a slight modification of the original algorithms. We then present the modification and theoretical analysis.

The modification is that we shall initially group the input links according to link diversity, and then choose the most weighted group of links as input of the two original algorithms. Let $g$ be a constant, and links with length in $ [g^{j-1}r, g^jr)$ belong to the same group $G_j$. Then we totally get $g(E)$ groups of links. Let $G_{j^*}$ be the most weighted group and input of Algorithm $2$ and $3$, then we have,
\begin{theorem}
    Algorithm 2 has an approximation ratio of $O(g(E))$; Algorithm 3 has an approximation of $O(g(E))$ when  $\rho$ is a constant and $O(g(E)\log\rho)$ otherwise.
\end{theorem}
\begin{IEEEproof}
    Please note that for links in $G_{j^*}$, the ratio between the longest links and shortest links becomes the constant $g$. The factor $1/\delta$ contained in previous results is then  replaced by $g$.
    We give the proof of Algorithm 2. Let $\mathcal{S}$ be the output, then,
    \[
        O(1)\cdot W(\mathcal{S}) \geq W(G_{j^*}).
    \]
Since $G_{j^*}$ is the most weighted group, we have,
\[W(G_{j^*}) \geq g(E)\cdot W(E) \geq g(E)\cdot W(OPT).\]
The proof for Algorithm 3 is similar.
\end{IEEEproof}
We also get an improved result under length-monotone, sub-linear fixed power assignments.
\begin{lemma}
Algorithm 3 achieves $O(g(E))$ approximation ratio for any length-monotone, sub-linear fixed power  assignment.
\end{lemma}

We then calculate some numeral results on these approximation ratios. Considering a typical wireless sensor network, we have $R=60$ and $r=5$. Let $\sigma = 3^\kappa, \kappa=3$. If we set $\alpha=2$, $g=2$ and use PTAS in the bridging process, then we have $\omega \approx 4^4$. For the adjustable power assignment, the ratio is around $1024 \cdot 12^{4\kappa+3}$. For the uniform power assignment, let $c^{up}=2$, and the approximation ratio is around $4^{3\kappa+7 }$. The computed approximation ratio for uniform power assignment is much better than adjustable power case. The larger approximation ratio for the adjustable power case is mainly caused by the constraint on a small value of $\phi^*$.

\subsection{Distributed implementation}
We then introduce how to implement distributed scheduling using our proposed algorithms.
Our previous works have developed  localized algorithms for the problem under the linear power setting  and uniform power setting \cite{my1}\cite{my-tpds-tr}.
The basic idea is that of partitioning the plane into super-subsquares of length $K$ cells, and performing centralized local scheduling in subsquares inside these super-subsquares.
Each subsquare has of length $(K-2M)$ cells--separating each local scheduling set, to limit the interference from other super-subsquares. The partitions are subsequently shifted so that all links can participate in the scheduling process.
To guarantee a globally feasible scheduling set consisting of all local scheduling set, we shall carefully set the distance of disjoint subsquares, i.e., the value of $M$. This could be derived using a similar method in \cite{my1}.
To provide theoretical guarantee for distributed implementation, we shall guarantee that the cardinality of these local scheduling set are  bounded by a constant from above. Obviously the local scheduling sets produced by algorithms in this paper satisfy this condition since the number of nodes in the scheduling set are bounded by a constant after an initial partition.

\section{Simulations}
In this section we evaluate performance of our proposed algorithms (Algorithm 2 and Algorithm 3) through simulation experiments. The throughput performance of scheduling algorithms is often measured by the total number of unscheduled packets, which is also termed the total backlog. Generally, the total backlog fluctuates slightly in a region if the arrival rate vector lies in the achievable capacity region of a link scheduling  algorithm. Inversely, the total backlog increases dramatically if the arrival rate vector exceeds the achievable capacity region. If the total backlog increases unboundedly to infinity, the network will become unstable.

In the following simulations, we will evaluate each algorithm in two network settings. One uses a randomly generated network topology and the other uses a real network topology from the CitySee project.
In the random network topology, we randomly select $20$ links as input from a network with $100$ nodes, half of which as senders randomly located on a plane with size $100\times100$ units, the other half as receivers positioned uniformly at random inside disks of radius $R=5$ around each of the senders. The minimum length of links is then set as $r=1$.
For the other setting, the network topology is part of the topology of the Citysee wireless sensor network, which is deployed for environment monitoring in the City Wuxi, China. The topology we use is shown in Fig. 1. (It uses the Cartesian coordinate system that is transformed from the geodetic coordinate system).
It contains $446$ nodes in an $1000m \times 1250m$ area.
The  maximum transmission range of the nodes outdoor is $100$ meters.
A link of such a large length is easy to fail in fixed power settings, thus we set the largest link length to be $60$  meters. We set the minimum  length of links as $10$ meters.

\begin{figure}[!htp]
\label{citysee}
\centering
\includegraphics[height=1.8in,width=3.5in]{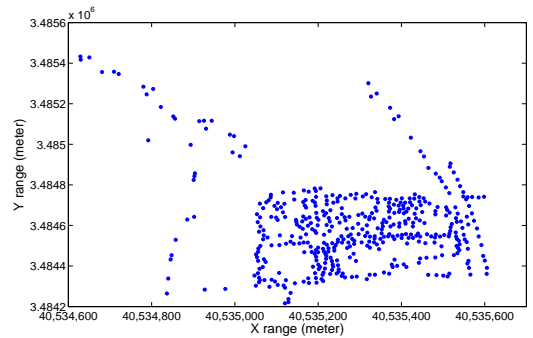}
\caption{Topology of the Citysee wireless sensor Network}
\label{fig2-nodes}
\end{figure}

Other common settings are as follows. The path loss exponent is set to be $3$ and the SINR threshold is $10$.
Packets arrive at each link independently according to a Poisson process with the same average arrival rate $\lambda$. Initially, we assign each link $k$ packets where $k$ is randomly chosen from $[100, 300]$.

\subsection{Algorithm 2 with adjustable power}
Next we present simulation results for Algorithm 2 with adjustable powers. We evaluate the throughput performance of Algorithm $2$, and verify correctness of the adjustable power assignment process.

We first present the throughput performance of Algorithm 2 under the random network topology.
We plot three figures to evaluate the maximum supportable average arrival rate in Fig. 2. We first study the fluctuation of the total backlog when the arrival rate increases from $0$ to the maximum link capacity of $1$. The increasing step is set to be $0.1$. It will give us an rough approximation of the achievable capacity region by link scheduling algorithms. Fig. 2(a) illustrates the trend of the total backlog at time slot $100000$ as the average arrival rate increases. Fig. 2(b)  zooms in the region of $[0.1,0.2]$ in Fig. 2(a).  It shows that the total backlog keeps stable around $0.185$. We then plot the fluctuation of the total backlog from time slot $0$ to time slot $100000$ under the average arrival rate $0.185$. In Fig. 2(c) it shows that the total backlog decreases rapidly  at the beginning, and then keeps stable in $[600,2000]$. It indicates that Algorithm 2 can still support an average arrival rate of $0.185$. Fig. 2(c) also illustrates the results when the average arrival rate is $0.190$, $0.195$ and $0.20$. The total backlog under $0.195$ still converges at a stable region, but it can not be stabilized under $0.20$. After an initial decrease, the total backlog for the average arrival rate $0.20$ increases nearly linearly since time slot $10000$. Thus we infer that Algorithm 2 can serve an maximum average arrival rate around $0.195$ under the random network topology.

Fig. 3 presents the assigned powers at different time slots for the random network setting. It respectively shows the maximum assigned power, the minimum assigned power and the average power per activated link at the selected time slots. The maximum assigned power is no greater than $20$, much smaller than the theoretical upper bound by Lemma $5$ ( The theoretical upper bound is $143$ in our setting). This verifies our theoretical analysis.

 \begin{figure*}[t]
    \centering
     \subfigure[Total backlog vs. average arrival rate]{
        \includegraphics[height=1.4in,width=2.1in]{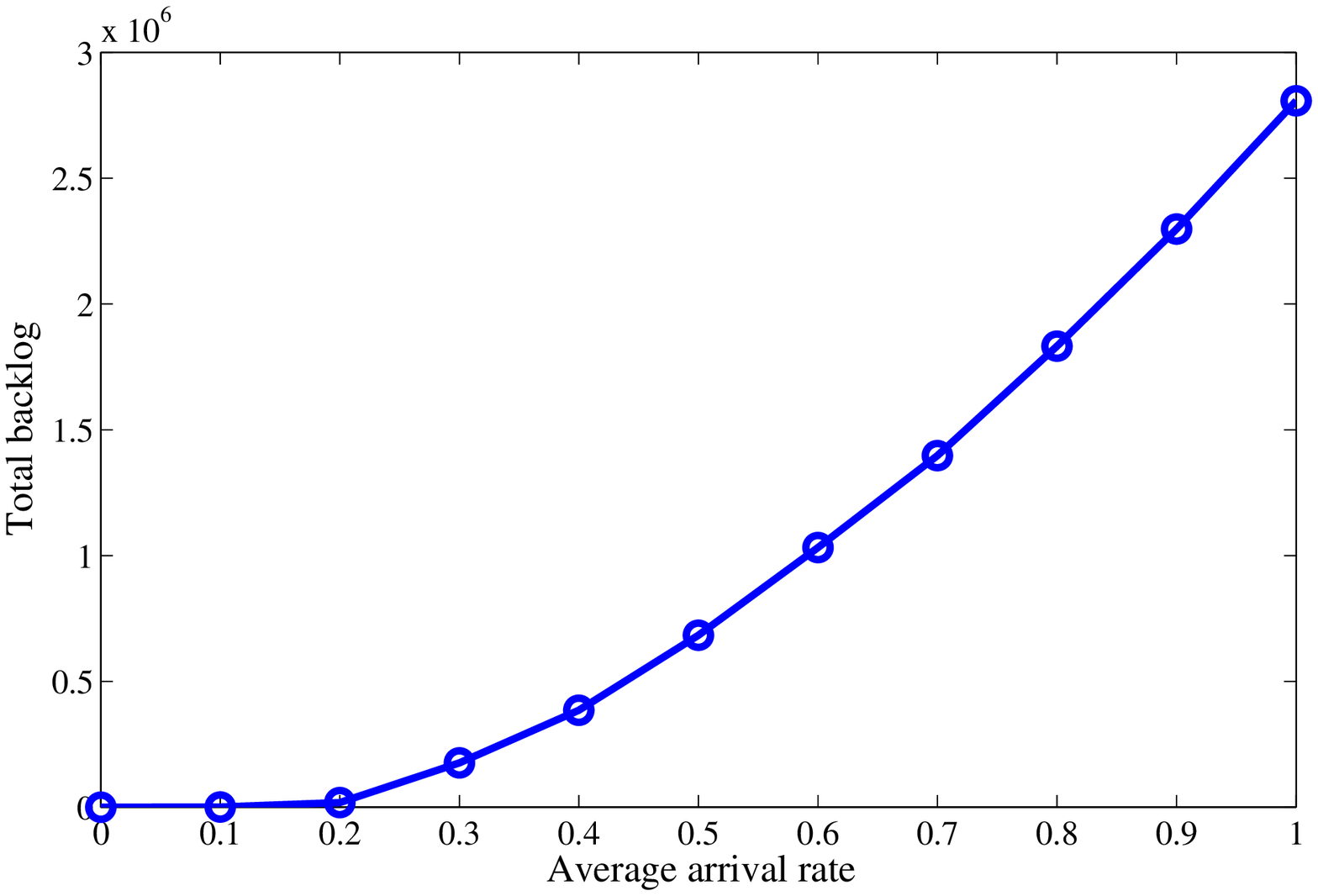}}
  \hspace{0cm}
    \subfigure[zoom in of (a)]{
        \includegraphics[height=1.4in,width=2.1in]{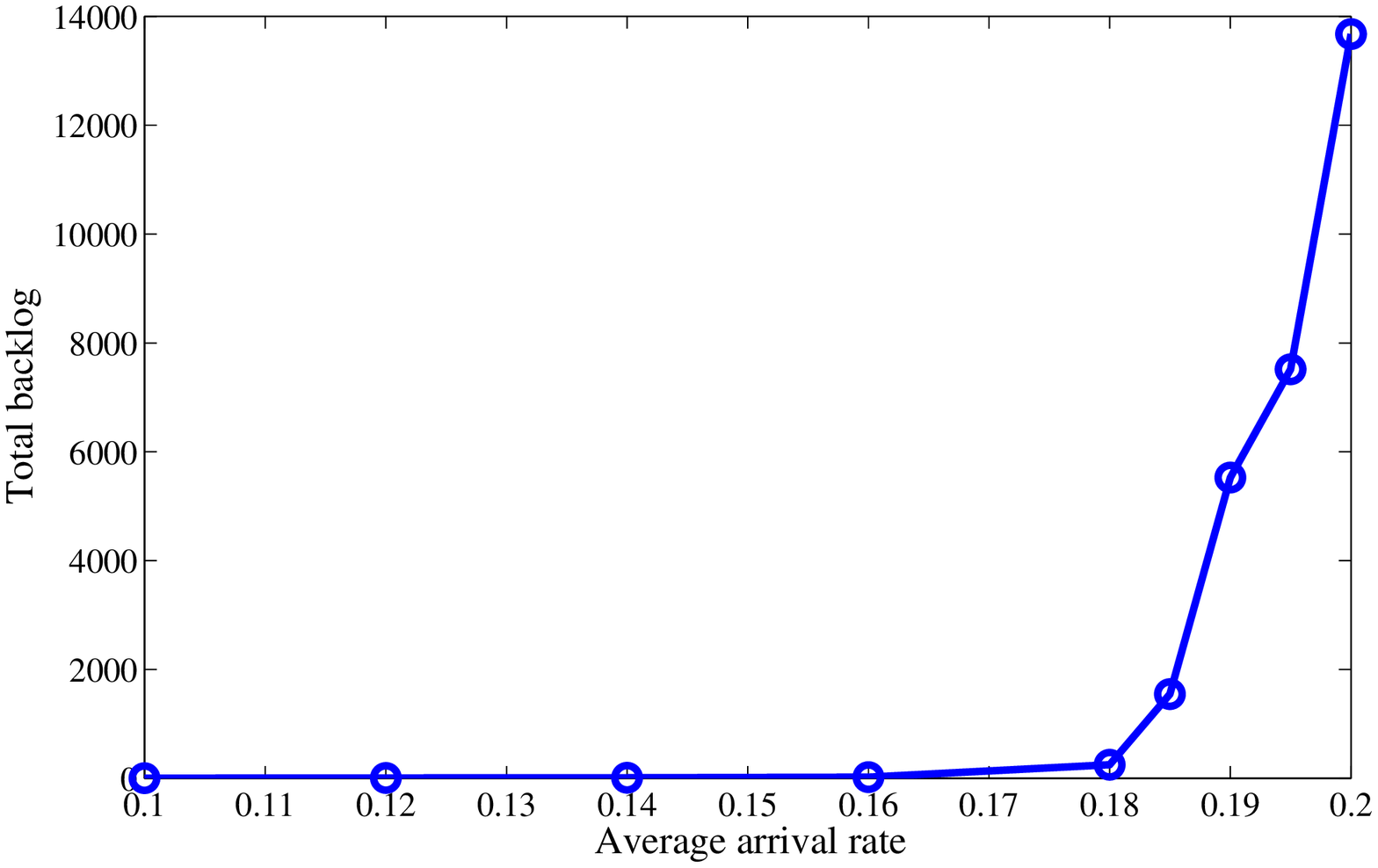}}
    \subfigure[Total backlog vs. time slot under different average arrival rate]{
        \includegraphics[height=1.4in,width=2.1in]{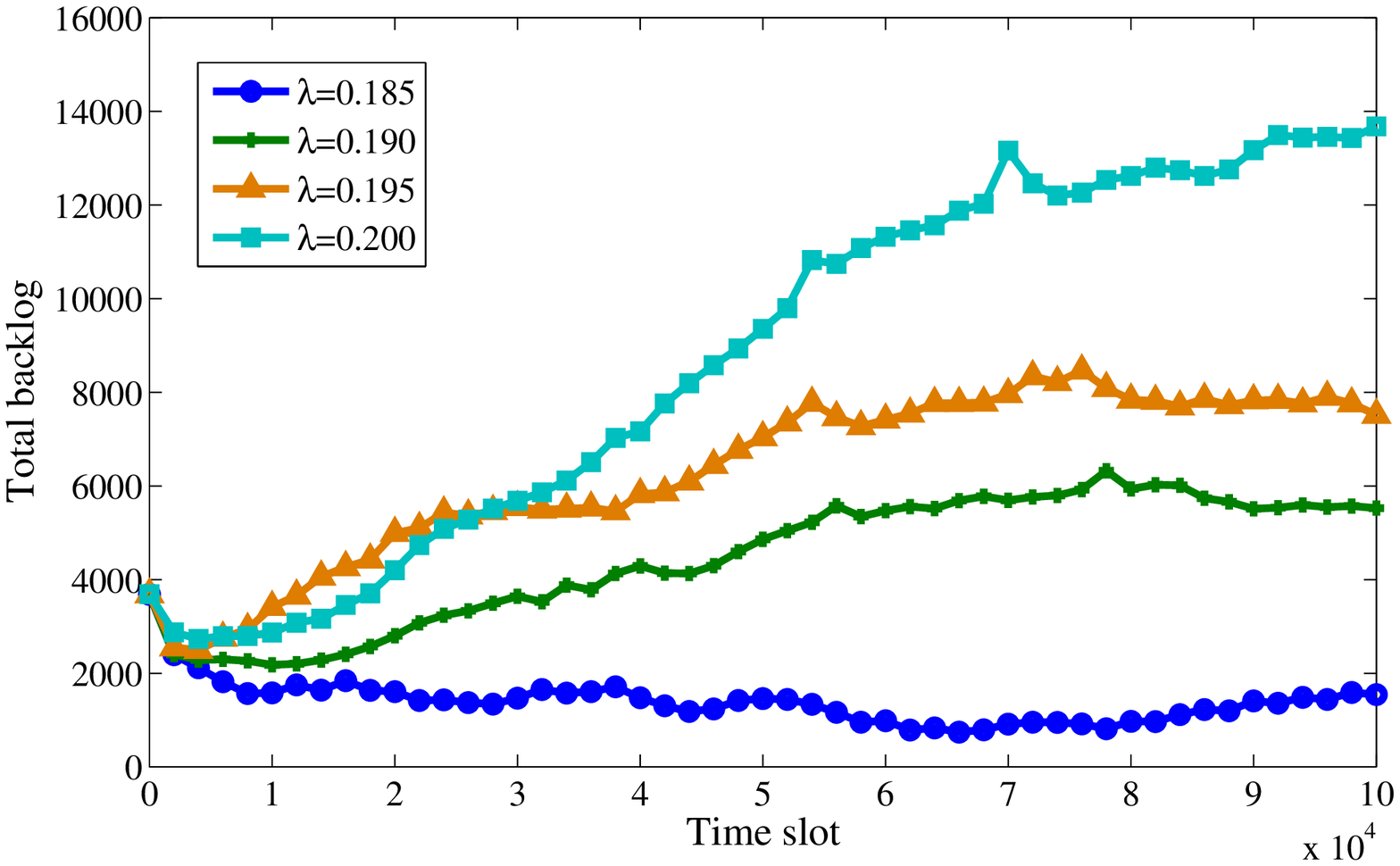}}
     \caption{Capacity region of Algorithm 2 under the random network topology }
 \end{figure*}

 \begin{figure}[!hbp]
    \centering
    \includegraphics[height=1.4in,width=2.6in]{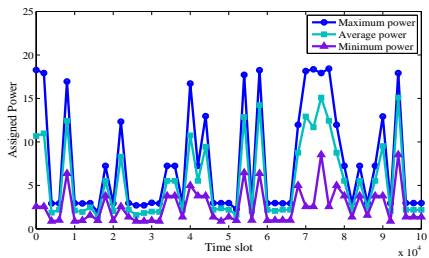}
    \caption{Power at different time slot under the random network topology}
 \end{figure}

We have also done the similar simulations and analysis for the Citysee network topology. The results on throughput performance and power assignments are shown in Fig. 4,
Fig. 5.
Similarly, combining the three subgraphs of Fig. 4, we can conclude that the maximum average arrival rate that Algorithm 2 achieves  is $0.01$ under the Citysee network setting. We then make some explanations that why the maximum average arrival rate takes such a low value.
According to the classical results in \cite{kumar2000},  an arbitrary wireless network can not provide an average throughput more than $O({1}/{\log|V|})$ if we use unit capacity. Thus we can roughly approximate that the optimal value is in the order of $0.047$ for the Citysee network. The comparison indicates that Algorithm 2 perform nearly optimally.

 \begin{figure*}[t]
    \centering
     \subfigure[Total backlog vs. average arrival rate]{
        \includegraphics[height=1.4in,width=2.1in]{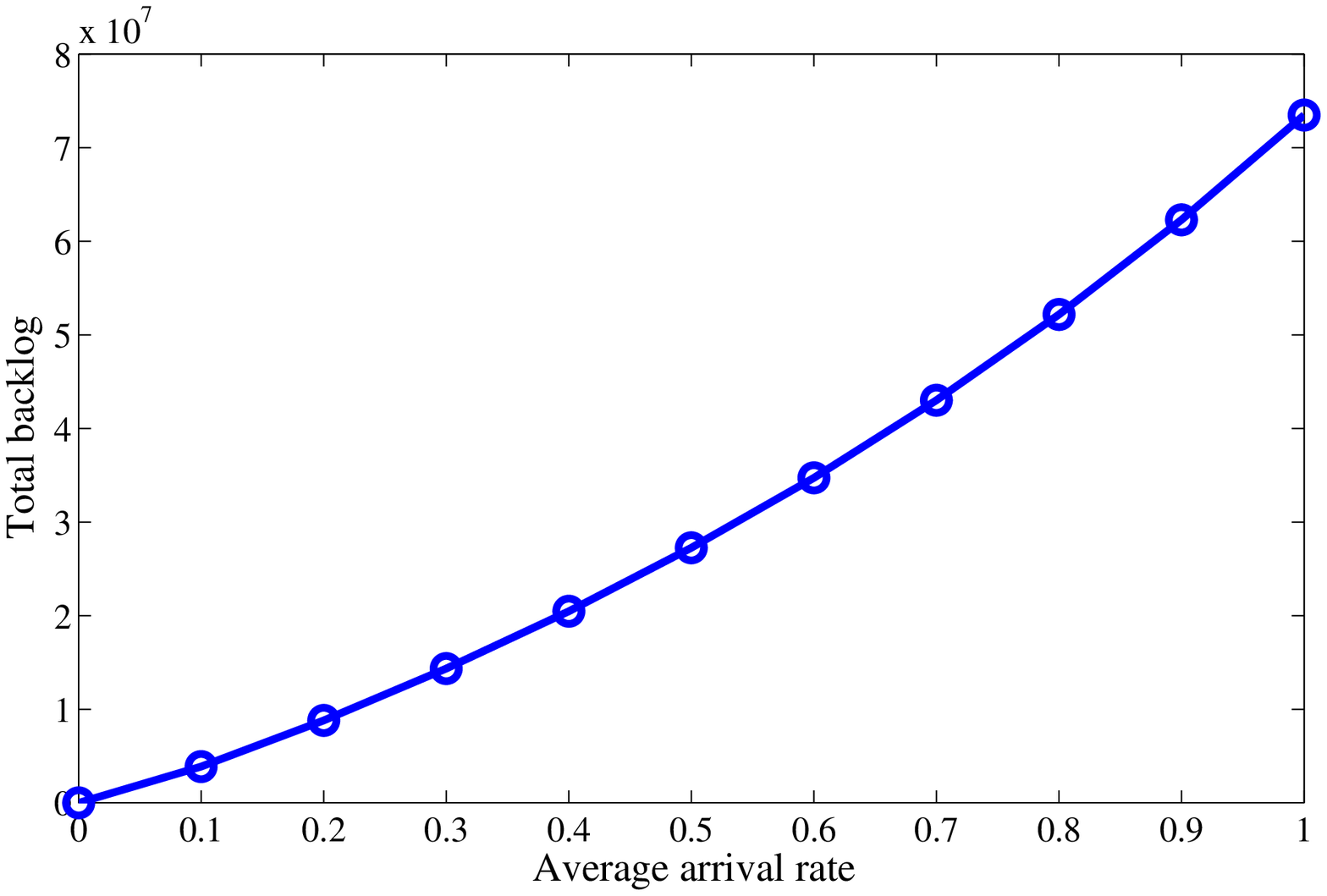}}
  \hspace{0cm}
    \subfigure[zoom in of (a)]{
        \includegraphics[height=1.4in,width=2.1in]{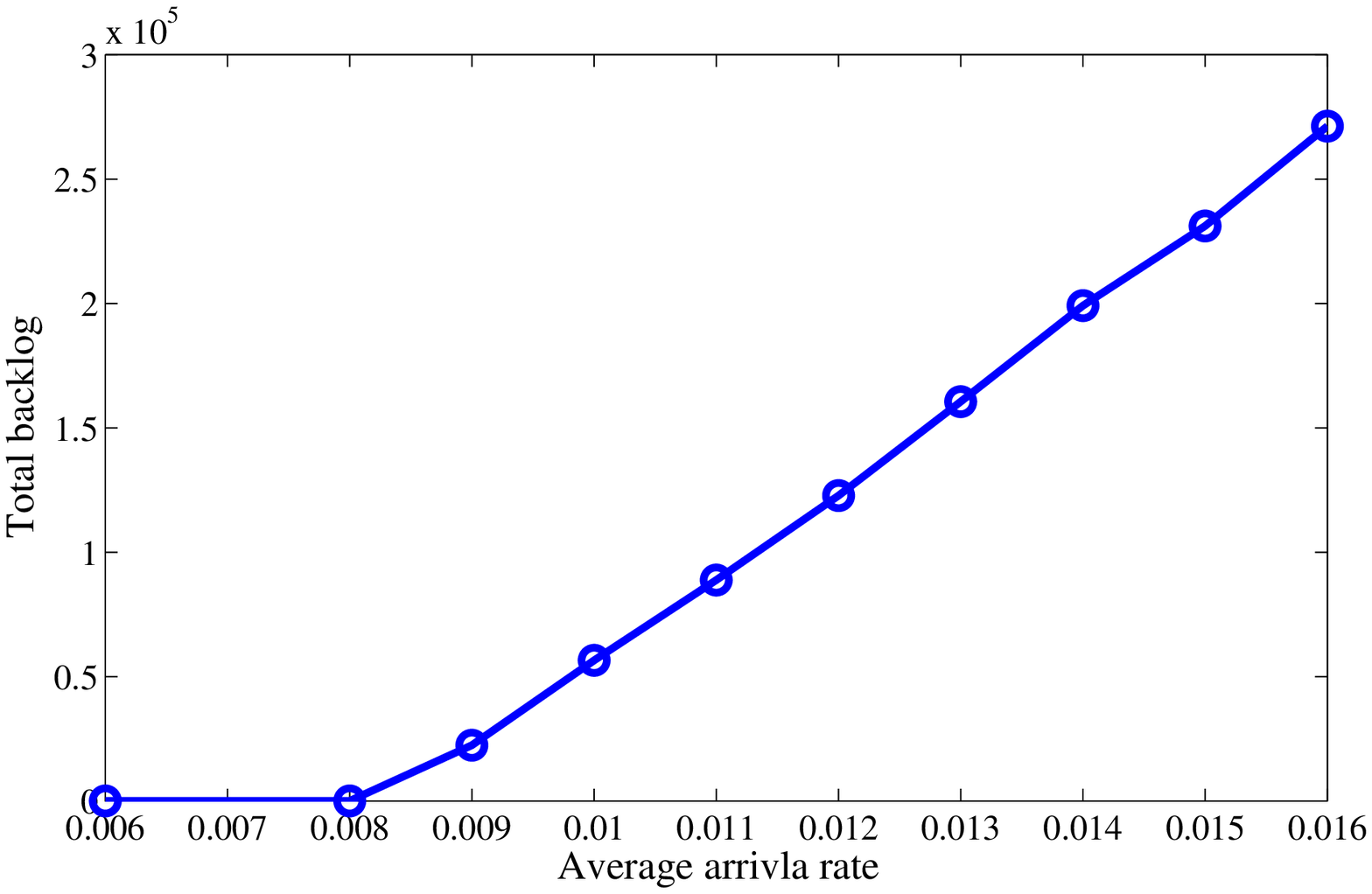}}
    \subfigure[Total backlog vs. time slot under different average arrival rate]{
        \includegraphics[height=1.4in,width=2.1in]{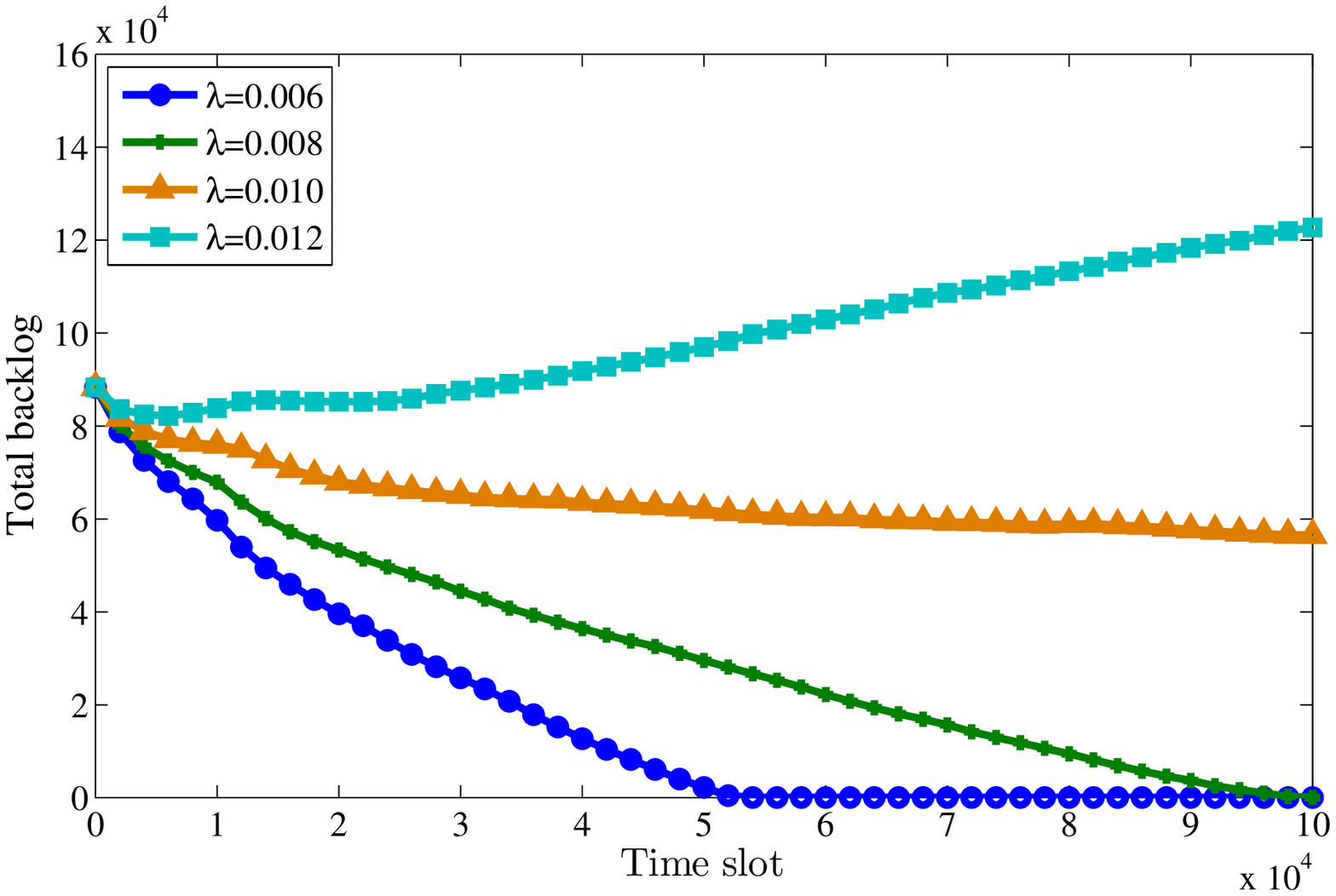}}
     \caption{Capacity region of Algorithm 2 under the Citysee topology }
 \end{figure*}

 \begin{figure}[!hbp]
    \centering
    \includegraphics[height=1.4in,width=2.6in]{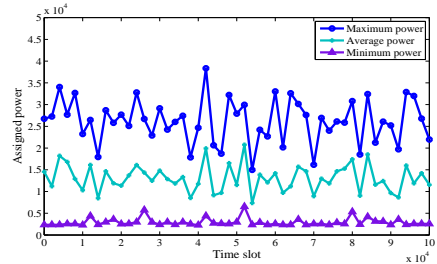}
    \caption{Power at different time slot under the Citysee topology}
 \end{figure}

\subsection{Algorithm 3 with fixed power}
 Fixed power assignments include lots of variants. It is difficult to conduct a comparative experiment for all fixed power settings, there being no obvious previous algorithms to compare it with. Here we focus on throughput performance under a commonly used uniform power assignment. By Theorem 2,  Algorithm 3 has best theoretical performance under the uniform power assignment because of the smallest power diversity among all fixed power assignments.
 We compare our algorithm with a logarithmic approximation algorithm \cite{S:phy9}, and the simple greedy maximal schedules\cite{S:phy7}  we can do comparison with it.

The algorithm in \cite{S:phy9} works as follows. First it removes the least weighted links. Next it partitions the remaining links into $\log(\Delta)$ groups according to their weights. $\Delta$ is the ratio of the maximum weight and the minimum weight among the remaining links. For each group, it finds a maximum independent set of links (MISL) by a constant-approximation algorithm. Then the most weighted MISL of the $\log(\Delta)$ MISLs is returned as the final result. We call this algorithm Weight for brevity.

The greedy algorithm works as follows. First it orders links in a decreasing order of weight. Going through the links, it  choose the most weighted link to the scheduling set. If the newly added link makes the scheduling set unfeasible, it will remove this link and turn to the next one. The process repeats until no links can be added. We refer to it as Greedy in the following paper.

\begin{figure}
    \centering
            \centering
                \subfigure[Total backlog vs. average arrival rate]{
                \includegraphics[height=1.2in,width=1.65in]{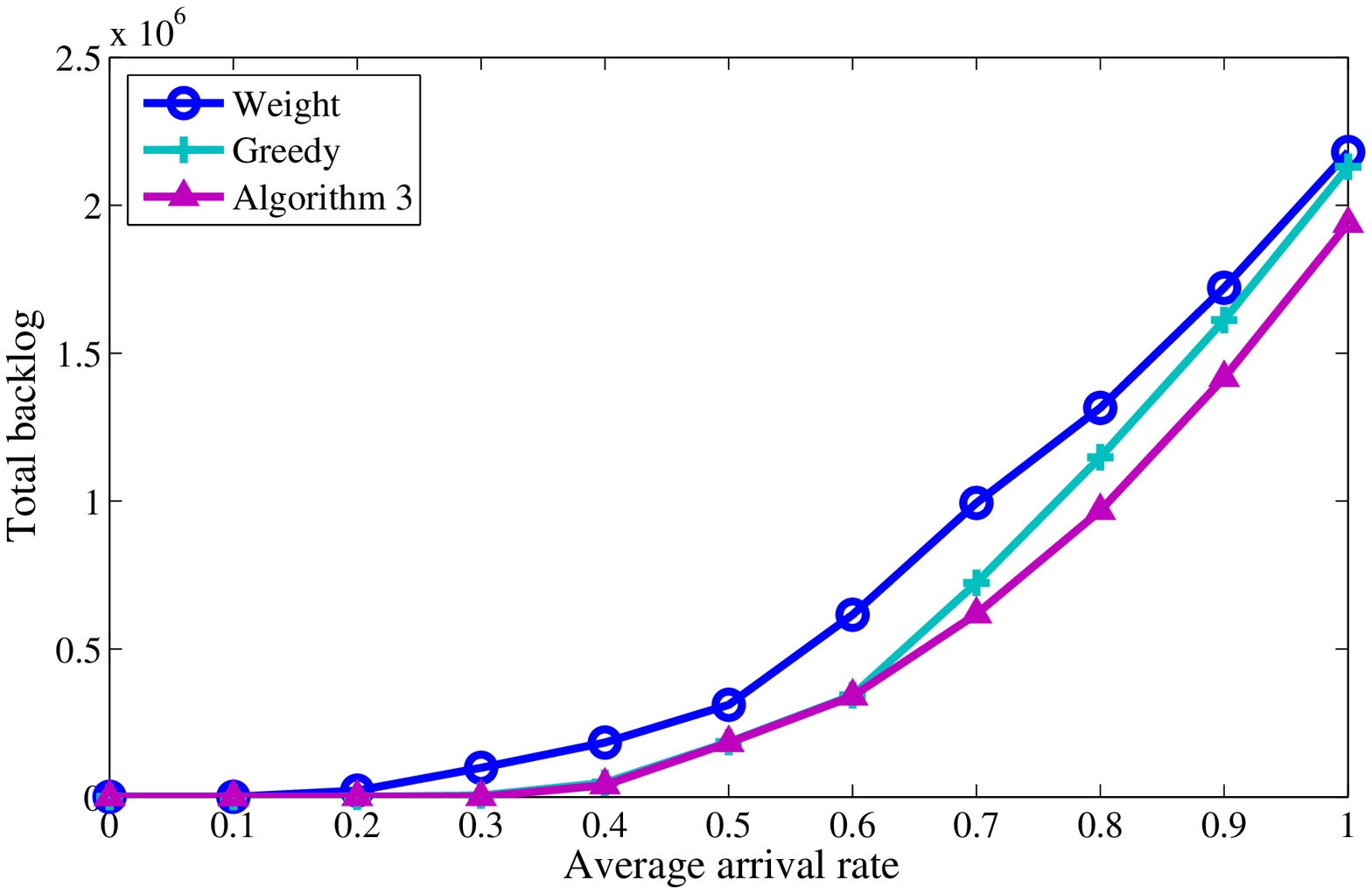}}
                \subfigure[zoom in of (a)]{
                \includegraphics[height=1.2in,width=1.65in]{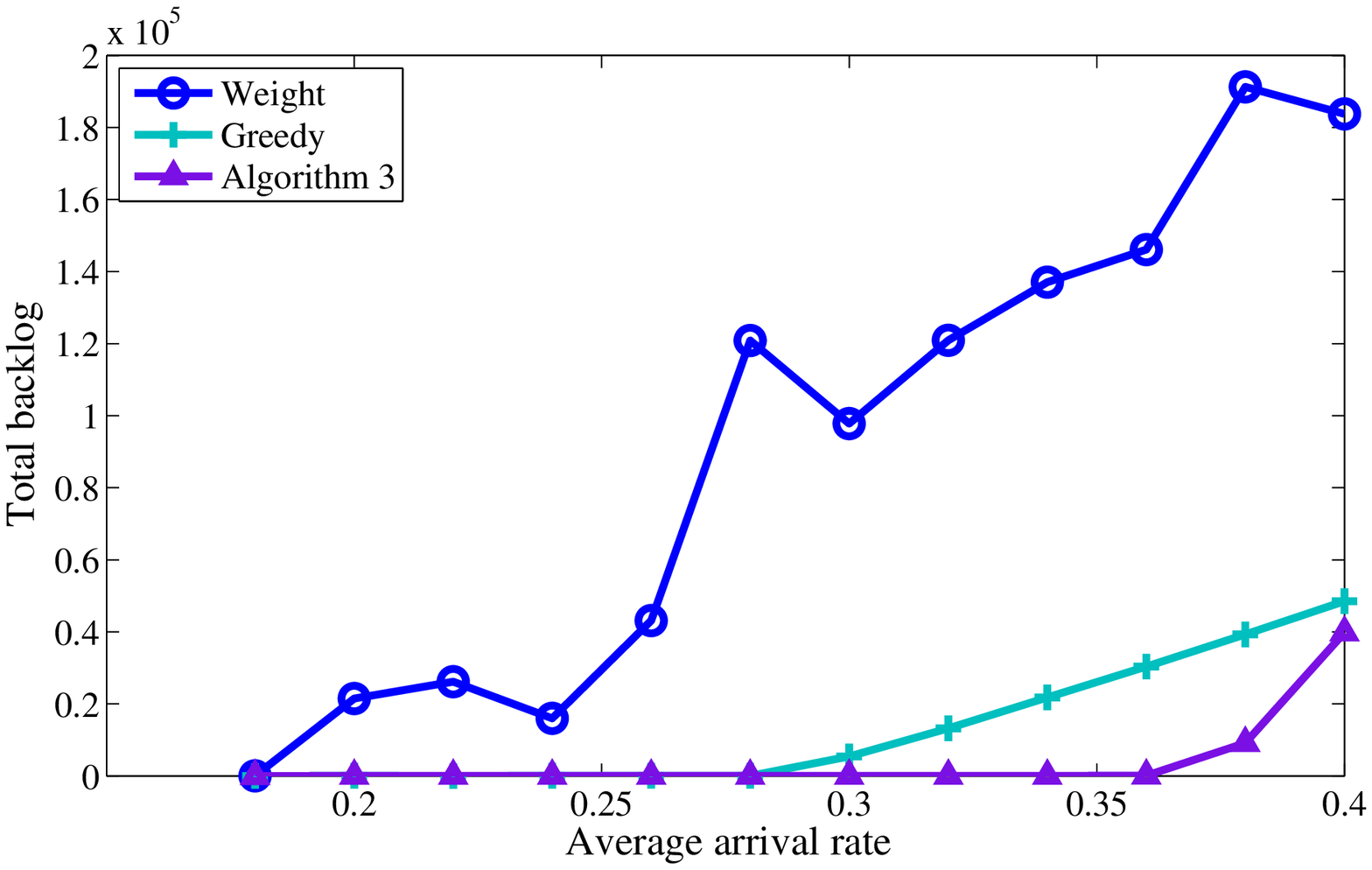}}
                \caption{Comparison between Weight, Greedy, and Algorithm 3  at different arrival rates under the random network topology }
\end{figure}

\begin{figure*}
            \centering
                   \subfigure[Weight]{
                    \includegraphics[height=1.4in,width=2.1in]{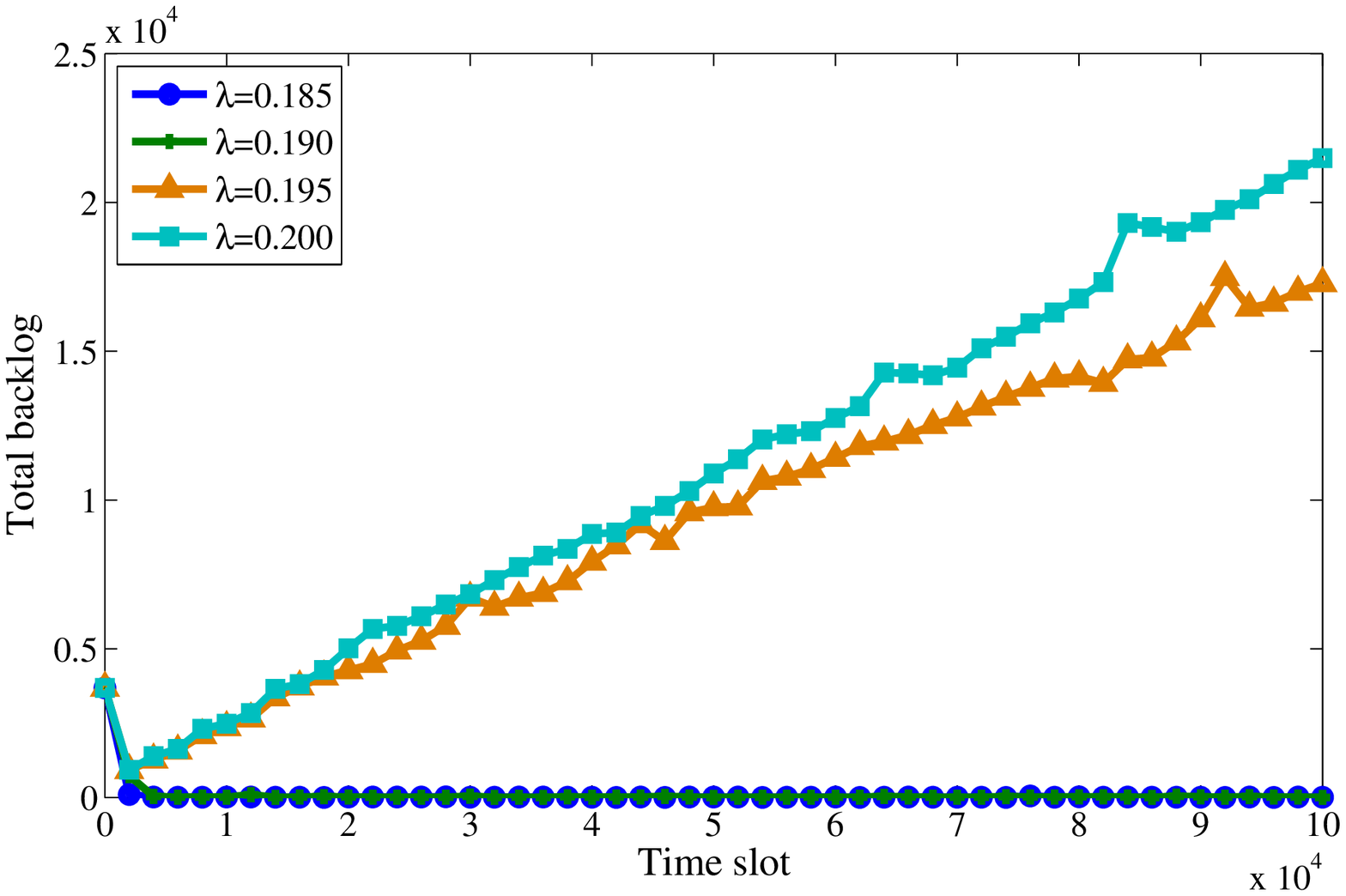}}
                    \subfigure[Greedy]{
                    \includegraphics[height=1.4in,width=2.1in]{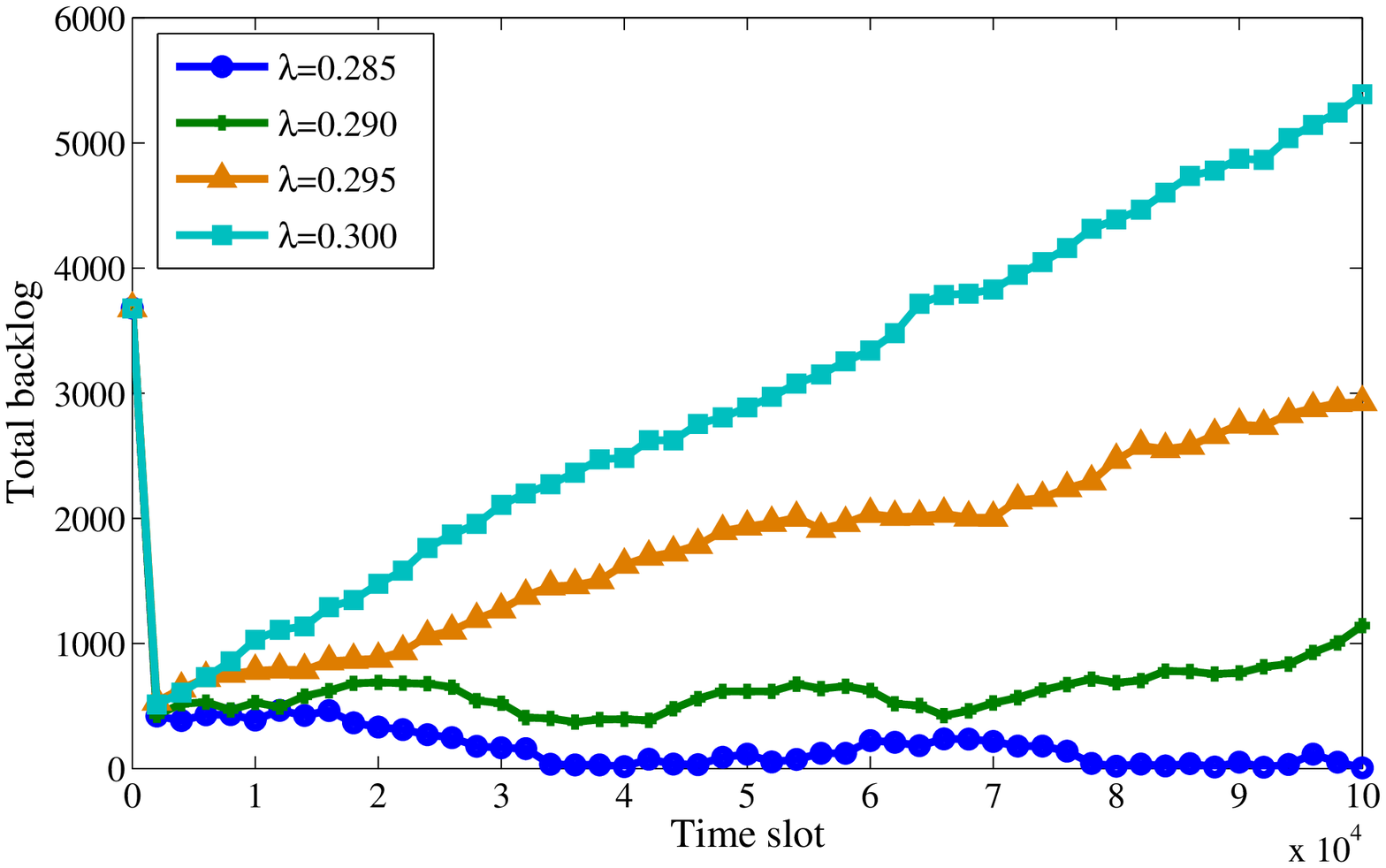}}
                    \subfigure[Algorithm 3]{
                    \includegraphics[height=1.4in,width=2.1in]{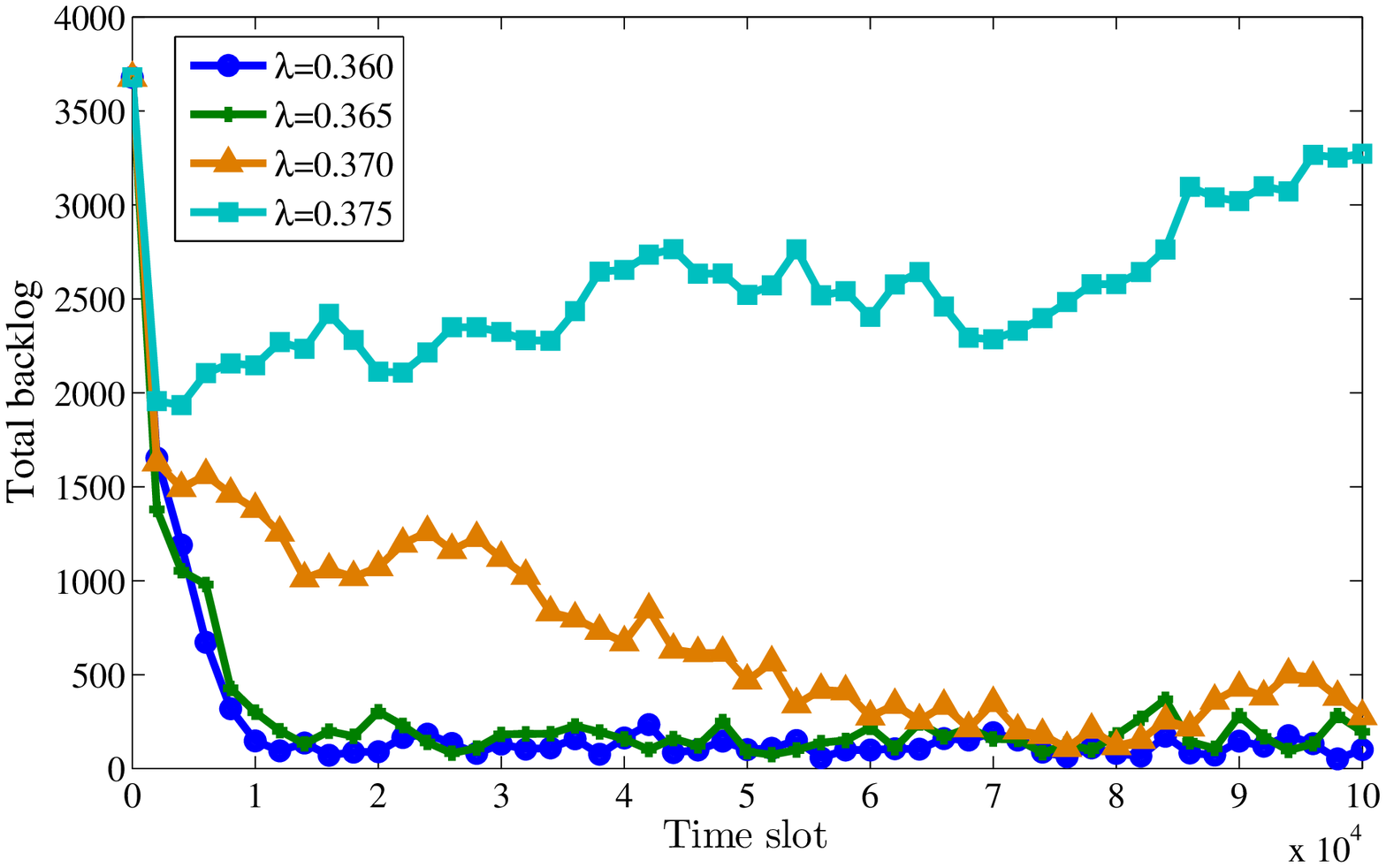}}
                    \caption{Achievable capacity region by Weight, Greedy, and Algorithm 3 under the random network topology }
\end{figure*}

 \begin{figure}
    \centering
            \centering
                \subfigure[Total backlog vs. average arrival rate]{
                    \includegraphics[height=1.2in,width=1.65in]{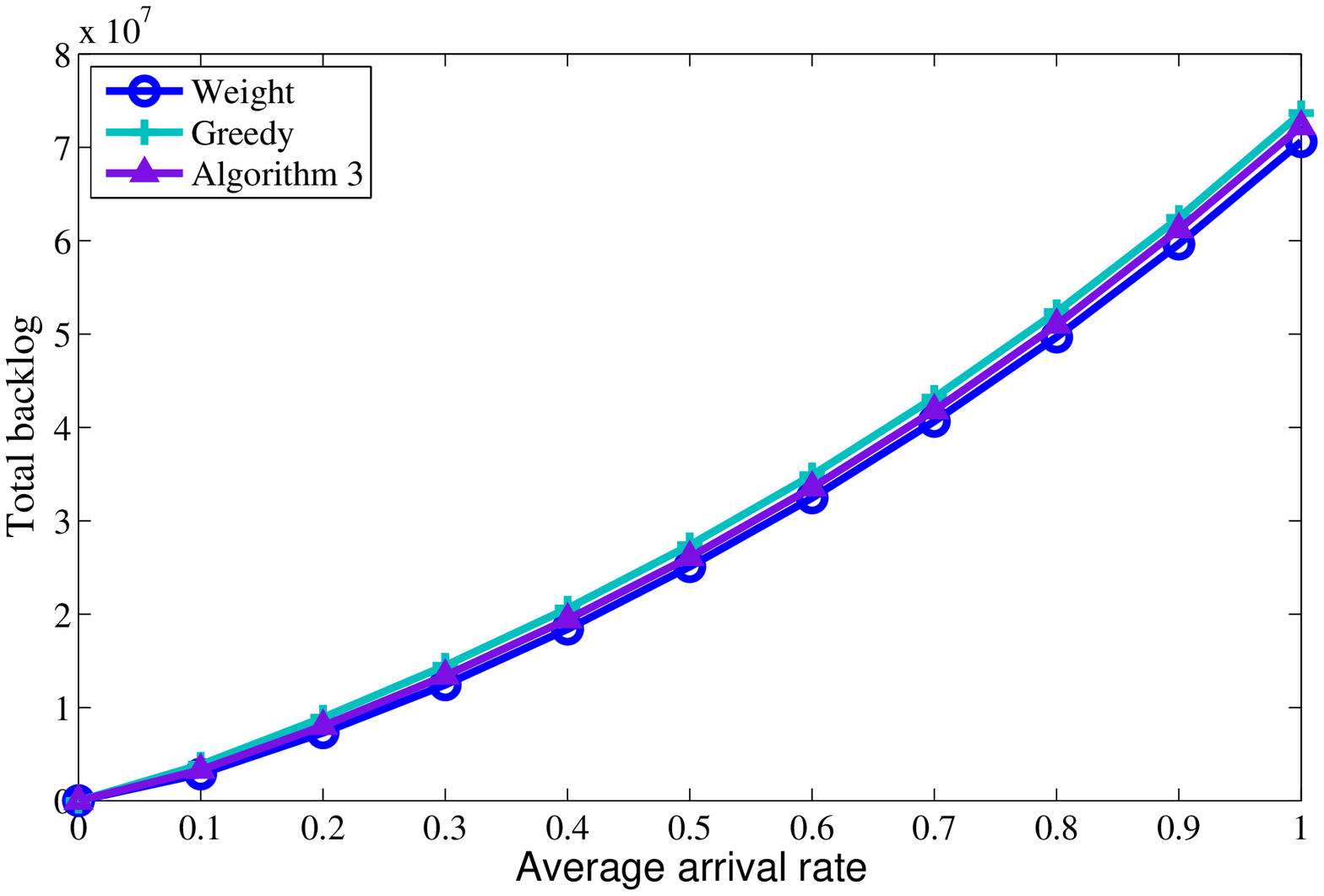}}
                \subfigure[zoom in of (a)]{
                     \includegraphics[height=1.2in,width=1.65in]{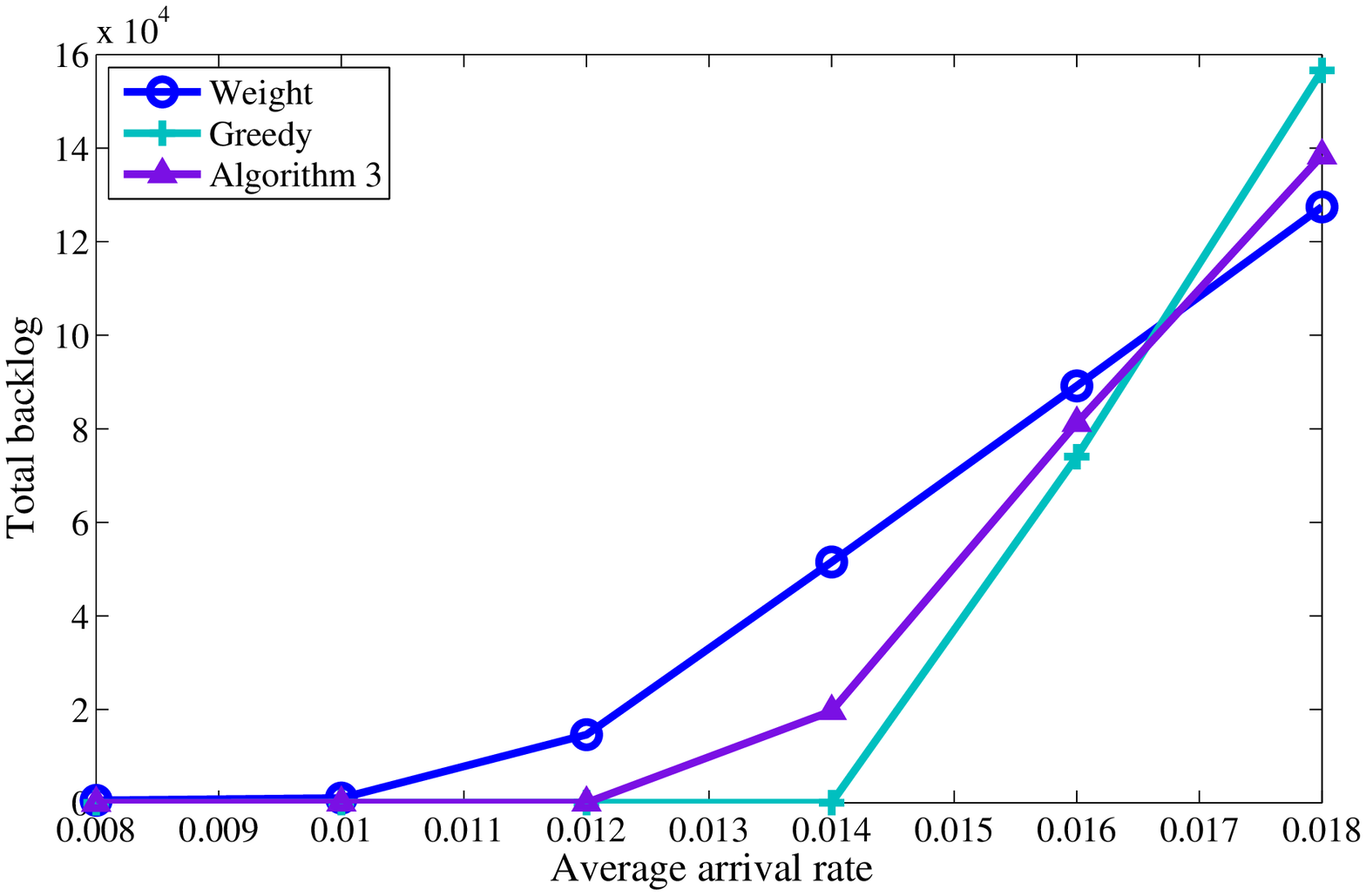}}
                 \caption{Comparison between Weight, Greedy, and Algorithm 3 at different arrival rates under the Citysee topology }
 \end{figure}

 \begin{figure*}
            \centering
                     \subfigure[Weight]{
                    \includegraphics[height=1.4in,width=2.1in]{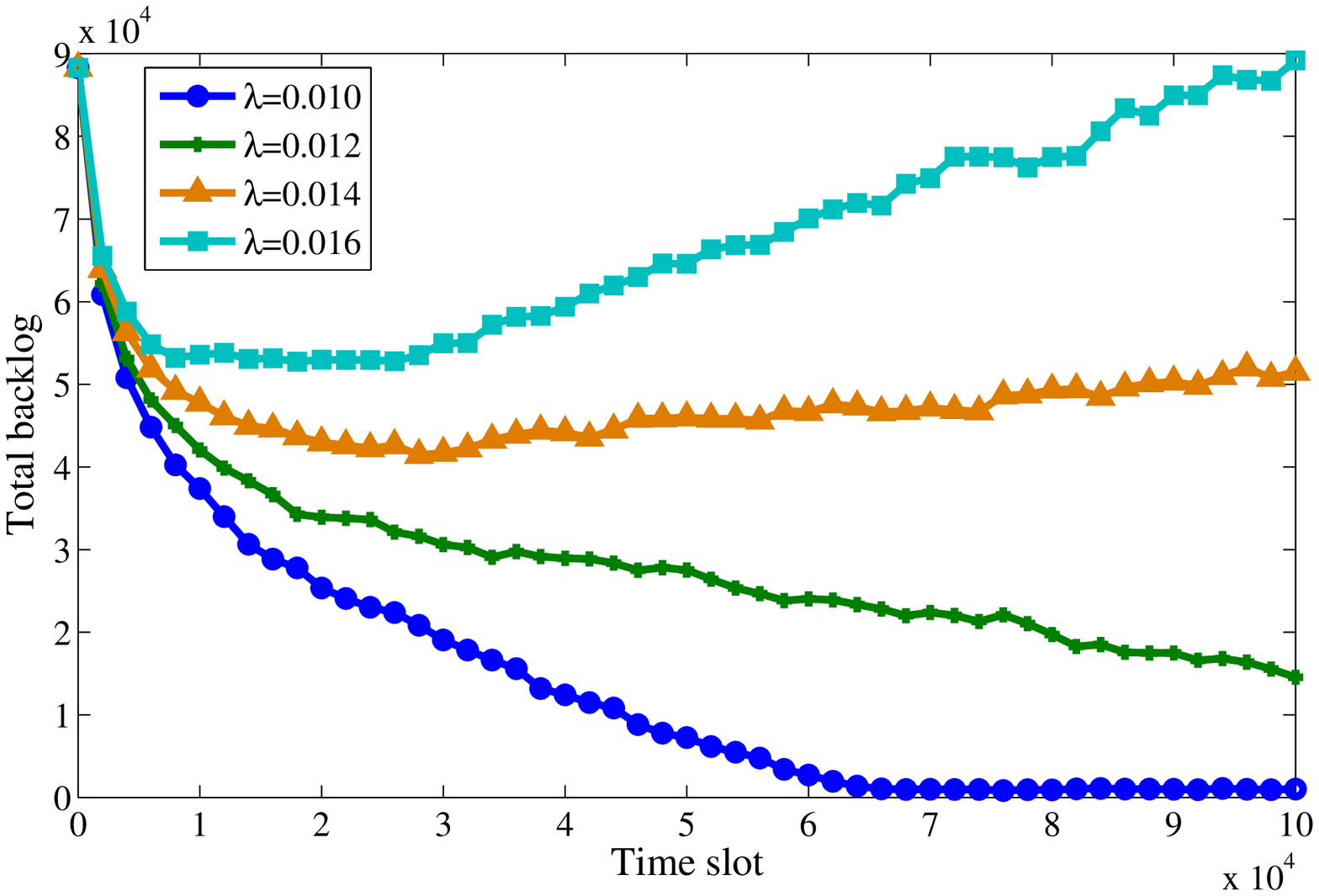}}
                    \subfigure[Greedy]{
                    \includegraphics[height=1.4in,width=2.1in]{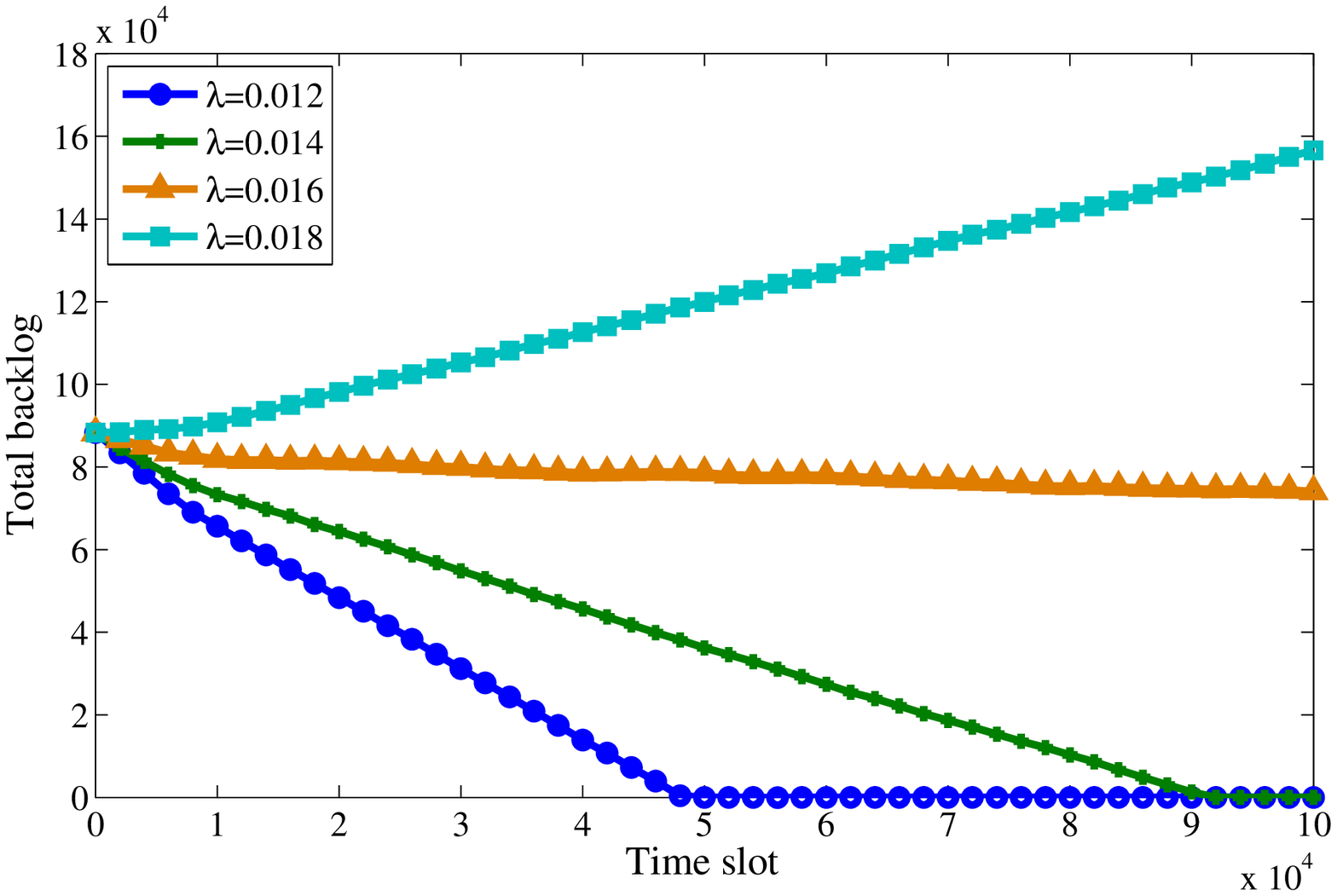}}
                    \subfigure[Algorithm 3]{
                    \includegraphics[height=1.4in,width=2.1in]{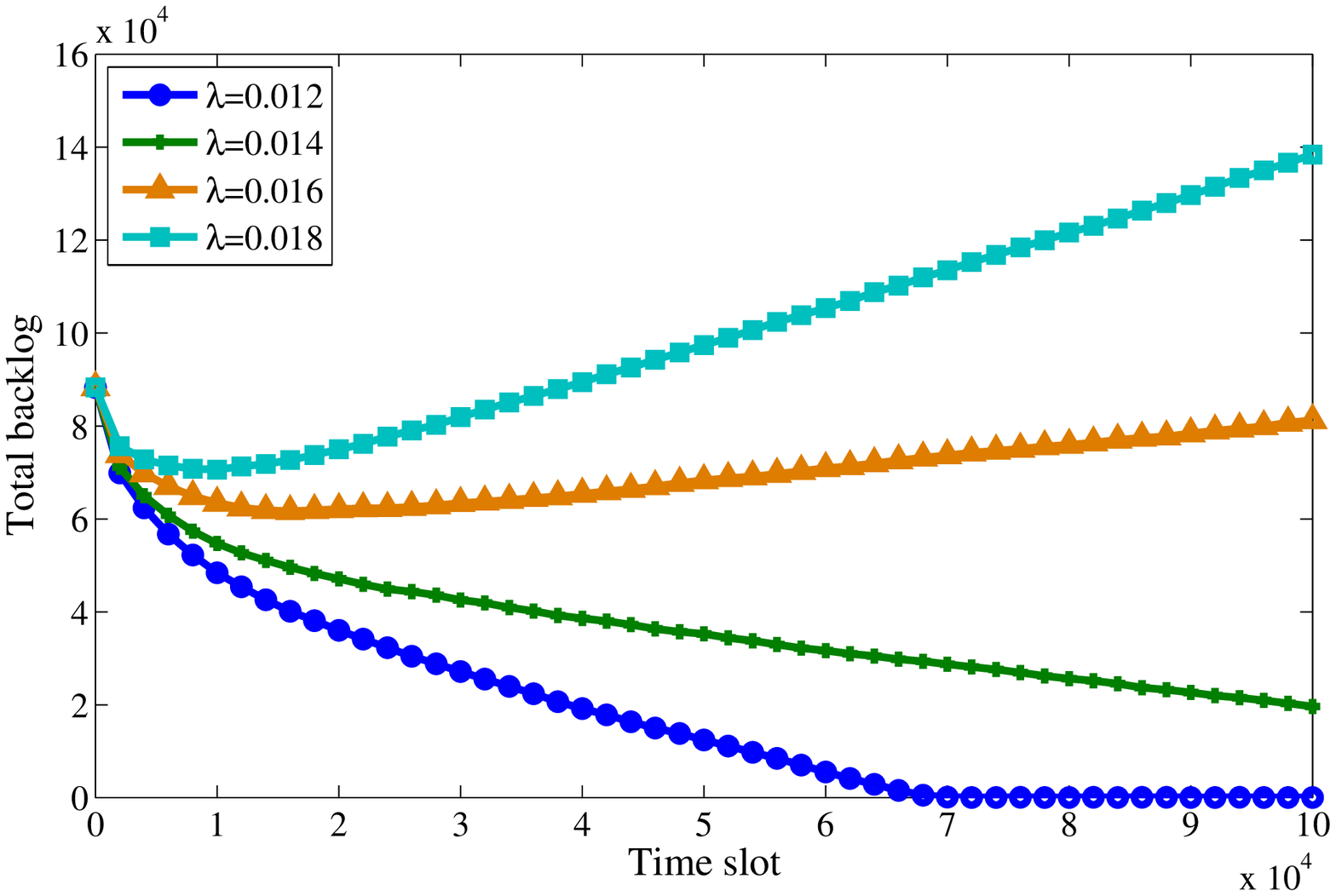}}
         \centering
        \caption{Achievable capacity region by Weight, Greedy, and Algorithm 3 under the Citysee topology }
 \end{figure*}

We set the same series of simulations as we do for Algorithm 2. The simulation results using the random network topology is shown in Fig. 6 and Fig. 7, and the simulation results using the Citysee network topology shown in Fig. 8 and Fig. 9. Algorithm 3 outperforms Weight and Greedy under the random  topology. Under the Citysee topology, it still has better performance than Weight, and the same performance as Greedy.

\section{Related works}
The link scheduling problem and its variants have been extensively studied in literature.
 Early works are mostly on graph-based models that simplify the complexity of wireless communication\cite{S:pick1}, \cite{S:pick2}, \cite{S:pick3}, \cite{S:constant1}, \cite{S:constant4}, \cite{S:GMS}, \cite{S:MWM2}, \cite{S:MS}. In the seminal work \cite{S:MWM1}, Tassiulas and Ephremides prove that the celebrated maximum weighted scheduling (MWS) achieves the optimal throughput capacity. Since finding a MWS is NP-hard in general interference models, a variety of simpler and/or suboptimal scheduling algorithms are proposed to achieve full or fractional optimal throughput capacity.

Under the physical interference model,  Chafekar \emph{et al.}  \cite{S:phy11} make a first attempt on logarithmic-approximation algorithms for the problem with the uniform and linear  power assignments. However, the attained bound is not relative to the original optimal throughput capacity, but to the optimal value by using slightly smaller power levels.  \cite{S:phy7} analyzes the performance of GMS under the physical model with uniform power assignment, and employs a technique named ``interference localization" to prevent the achievable performance vanishing.
Xu \emph{et al.} \cite{S:phy5} firstly  get a constant-approximation algorithm for the MWISL problem with linear power assignment. A subsequent work gains a logarithmic-approximation factor related to ratio between the maximum and minimum weight for the uniform case\cite{S:phy9}.
Most recently, Halld\'{o}rsson and Mitra  also claim a constant-approximation ratio for the linear power setting, and  poly-logarithmic approximation ratios dependent on size of link set for other  length-monotone, sub-linear fixed power settings \cite{halldosson2012infocom}. The proposed algorithms utilize a LP based approach to find a link set with constant affectance, and then refine the set into a feasible scheduling set. Nevertheless, they have to rely upon a huge constant (the exact value is not specified in \cite{halldosson2012infocom}) to upper bound the affectance, which results in a quite small approximation ratio in the order of the square of the huge constant.

All these aforementioned algorithms are centralized, some works also develop distributed link scheduling algorithms for practical applications. Zhou \emph{et al.} \cite{my1} firstly propose a distributed algorithm with a constant-approximation ratio for the linear power case, and  a randomized vision is also seen in \cite{Halldorsson2012ciss}. \cite{Pei2012mobihoc} very recently proposed a low complexity scheduling algorithm for a special fixed power assignment where transmission powers of two links with almost equal length are within a constant from each other.
\cite{S:dphy1} proposes a CSMA-type distributed link scheduling approach with throughput optimality for the uniform power case. However, this approach has high communication overhead

A quite related work \cite{modiano2009wiopt} studies the distributed throughput maximization problem via random power allocation under the SINR-RATE based interference model. In such a interference model, the capacity of a link is not a fixed value(e.g., $1$ if SINR threshold satisfied and $0$ otherwise), but determined by the SINR value at the receiver (i.e., $\log(1+SINR_i)$).
 For simplification, it assumes  static path  gain over time, whereas the gain is actually determined by concurrent transmissions and thus varies over time.  Consequently, the problem studied in  \cite{modiano2009wiopt} does not  include an ISL problem with complex interference constraints. The solution bases on  a pick-and-compare approach\cite{S:pick3} to asymptotically achieve the optimal. However, the probability of this near-optimal approach is quite low (i.e., the probability is $4N^{-N}$ where $N$ is the number of nodes). The simulation results in  \cite{modiano2009wiopt}  show that it can just stabilize an arrive rate of 0.03 under a random network of 16 nodes, while we can support an maximum arrival rate of $0.37$ under a random network of 20 Nodes.

Two related problems on capacity are the capacity maximization problem which seeks a maximum number of independent links of a given set of links, and the minimum length scheduling problem which seeks a partition of a given set of links into the fewest independent sets. We make a brief review on the problems under the context of physical interference.

For the capacity maximization problem, \cite{Halldorsson2011soda} and \cite{Kesselheim2011soda} respectively achieve a constant-approximation factor with the oblivious power and power control. However, to ignore the influence of ambient noise, \cite{Kesselheim2011soda} has to assume arbitrary transmission power for each link. This assumption is not reasonable in practice. Motivated by this, Wan \emph{et al.} \cite{wan2012infocom} then get a constant-approximation algorithm which does not assume unbounded maximum transmission power. A distributed implementation with a constant-approximation factor is proposed in \cite{Akummar2012} which implicitly assumes the uniform power assignment. The algorithm makes a strong  assumption that all nodes have physical carrier sensing capability and can detect if the sensed signal exceeds a threshold. This assumption undoubtedly reduces the difficulties because the main challenge of the original problem is to locally approximate and bound the unknown global interference.

For the minimum length scheduling problem, the overall state-of-the-art  retains in the order of logarithm under the uniform power setting \cite{S:phy8} \cite{S:phy6} \cite{S:phy17}. In \cite{S:phy8}, an attempt on a constant-approximation algorithm for this problem with uniform power assignment fails, and the claim has been retracted by the authors recently.

\section{Conclusion}
We tackle the link scheduling problem for throughput  maximization under the physical interference model.
We solve two variants of the problem by developing approximation algorithms for MWISL problem in a unified scheme.
Our algorithms are based on our discovery of intrinsic connections between the SINR-based and graph-based interference. Our results are applicable to the minimum  length scheduling problem and the maximum multiflow problem from an algorithmic reduction view\cite{S:phy6}.

Many  problems remain open and are left for future works. Our current approximation ratios are related to link diversity and power diversity. It is still open that whether there exists constant approximation independent of these network parameters.
  Meanwhile, these results in this work are proved to hold in a special fading metric space (the Euclidean plane). It is unknown whether the same results are attainable in general metric spaces.
 Moreover, all aforementioned challenges are limited to the objective of long-term throughput maximization. Other
  SINR-constrained link scheduling problem with different optimization objectives, or effective multihop flow scheduling with these optimization objectives, still need better solutions.

\bibliographystyle{IEEEtran}
\bibliography{mylib}

\begin{thebibliography}{10}
\providecommand{\url}[1]{#1}
\csname url@samestyle\endcsname
\providecommand{\newblock}{\relax}
\providecommand{\bibinfo}[2]{#2}
\providecommand{\BIBentrySTDinterwordspacing}{\spaceskip=0pt\relax}
\providecommand{\BIBentryALTinterwordstretchfactor}{4}
\providecommand{\BIBentryALTinterwordspacing}{\spaceskip=\fontdimen2\font plus
\BIBentryALTinterwordstretchfactor\fontdimen3\font minus
  \fontdimen4\font\relax}
\providecommand{\BIBforeignlanguage}[2]{{%
\expandafter\ifx\csname l@#1\endcsname\relax
\typeout{** WARNING: IEEEtran.bst: No hyphenation pattern has been}%
\typeout{** loaded for the language `#1'. Using the pattern for}%
\typeout{** the default language instead.}%
\else
\language=\csname l@#1\endcsname
\fi
#2}}
\providecommand{\BIBdecl}{\relax}
\BIBdecl

\bibitem{S:ptas}
X.-Y. Li and Y.~Wang, ``Simple approximation algorithms and {PTASs} for various
  problems in wireless ad hoc networks,'' \emph{Journal of Parallel and
  Distributed Computing}, vol. $66$, pp. 515--530, 2006.

\bibitem{sharma2006complexity}
G.~Sharma, R.~Mazumdar, and N.~Shroff, ``On the complexity of scheduling in
  wireless networks,'' in \emph{Proc. ACM MobiCom}, 2006, pp. 227--238.

\bibitem{S:phy8}
M.~Halld\'{o}rsson and R.~Wattenhofer, ``Wireless communication is in {APX},''
  in \emph{Proc. 36th International Colloquium on Automata, Languages and
  Programming}, 2009, pp. 525--536.

\bibitem{S:phy7}
L.-B. Le, E.~Modiano, C.~Joo, and N.~B. Shroff, ``Longest-queue-first
  scheduling under {SINR} interference model,'' in \emph{Proc. ACM Mobihoc},
  2010, pp. 41--50.

\bibitem{S:MWM1}
L.~Tassiulas and A.~Ephremides, ``Stability properties of constrained queueing
  systems and scheduling policies for maximum throughput in multihop radio
  networks,'' \emph{IEEE/ACM Transactions on Automatic Control}, vol. $37$, pp.
  1936--1948, Dec. 1992.

\bibitem{S:phy5}
X.-H. Xu, S.-J. Tang, and P.-J. Wan, ``Maximum weighted independent set of
  links under physical interference model,'' in \emph{LNCS}, vol. $6221$.\hskip
  1em plus 0.5em minus 0.4em\relax Springer, Heidelberg, 2010, pp. 68--74.

\bibitem{halldosson2012infocom}
M.~M. Halld\'{o}rsson and P.~Mitra, ``Wireless capacity and admission control
  in cognitive radio,'' in \emph{Proc. IEEE Infocom}, 2012, pp. 855--863.

\bibitem{my1}
Y.~Q. Zhou, X.-Y. Li, M.~Liu, Z.~C. Li, S.~J. Tang, X.~F. Mao, and Q.~Y. Huang,
  ``Distributed link scheduling for throughput maximization under physical
  interference model,'' in \emph{Proc. IEEE Infocom}, 2012, pp. 2691--2695.

\bibitem{S:phy9}
X.~H. Xu, S.~J. Tang, and X.-Y. Li, \emph{Stable Wireless Link Scheduling
  Subject to Physical Interferences With Power Control}, 2011, manuscript.

\bibitem{S:phy6}
P.-J. Wan, O.~Frieder, X.-H. Jia, F.~Yao, X.-H. Xu, and S.-J. Tang, ``Wireless
  link scheduling under physical interference model,'' in \emph{Proc. IEEE
  Infocom}, 2011, pp. 838--845.

\bibitem{Pei2012mobihoc}
G.~Pei and A.~Vullikanti, ``Low-complexity scheduling for wireless networks,''
  in \emph{Proc. ACM MobiHoc '12}, 2012, pp. 35--44.

\bibitem{Neely2006ftn}
L.~Georgiadis, M.~J. Neely, and L.~Tassiulas, ``Resource allocation and
  cross-layer control in wireless networks,'' \emph{Found. Trends Netw.},
  vol.~1, no.~1, pp. 1--144, Apr. 2006.

\bibitem{halldorsson_2009}
M.~Halld¨®rsson, ``Wireless scheduling with power control,'' in
  \emph{Algorithms - ESA}, 2009, vol. 5757, pp. 361--372.

\bibitem{S:phy1}
O.~Goussevskaia, Y.~Oswald, and R.~Wattenhofer, ``Complexity in geometric
  {SINR},'' in \emph{Proc. ACM Mobihoc}, 2007, pp. 100--109.

\bibitem{wan2012infocom}
P.-J. Wan, D.~C. Chen, G.~J. Dai, Z.~Wang, and F.~Yao, ``Maximizing capacity
  with power control under physical interference model in duplex mode,'' in
  \emph{Proc. IEEE Infocom}, 2012, pp. 415--423.

\bibitem{S:pick3}
E.~Modiano, D.~Shah, and G.~Zussman, ``Maximizing throughput in wireless
  networks via gossiping,'' in \emph{Proc. ACM SIGMETRICS}, 2006, pp. 27--38.

\bibitem{S:GMS}
C.~Joo, X.~Lin, and N.~B. Shroff, ``Understanding the capacity region of the
  greedy maximal scheduling algorithm in multi-hop wireless networks,'' in
  \emph{Proc. IEEE Infocom}, 2008, pp. 1103--1111.

\bibitem{lin2005info}
X.~Lin and N.~B. Shroff, ``The impact of imperfect scheduling on cross-layer
  rate control in multihop wireless networks,'' in \emph{Proc. IEEE Infocom},
  2005, pp. 1804--1814.

\bibitem{Kesselheim2011soda}
T.~Kesselheim, ``A constant-factor approximation for wireless capacity
  maximization with power control in the {SINR} model,'' in \emph{Proceedings
  of the Twenty-Second Annual ACM-SIAM Symposium on Discrete Algorithms}, 2011,
  pp. 1549--1559.

\bibitem{wan2011maximizing}
P.~Wan, C.~Ma, S.~Tang, and B.~Xu, ``Maximizing capacity with power control
  under physical interference model in simplex mode,'' \emph{Wireless
  Algorithms, Systems, and Applications}, pp. 84--95, 2011.

\bibitem{my-tpds-tr}
Y.~Q. Zhou, X.-Y. Li, M.~Liu, X.~F. Mao, S.~J. Tang, and Z.~C. Li,
  ``{Throughput Optimizing Localized Link Scheduling for Multihop Wireless
  Networks Under Physical Interference Model},'' \emph{ArXiv e-prints,
  http://arxiv.org/abs/1301.4738}, Jan. 2013.

\bibitem{kumar2000}
P.~Gupta and P.~Kumar, ``The capacity of wireless networks,'' \emph{Information
  Theory, IEEE Transactions on}, vol.~46, no.~2, pp. 388 --404, mar 2000.

\bibitem{S:pick1}
S.~Sanghavi, L.~Bui, and R.~Srikant, ``Distributed link scheduling with
  constant overhead,'' in \emph{Proc. ACM SIGMETRICS}, 2007, pp. 313--324.

\bibitem{S:pick2}
S.-J. Tang, X.-Y. Li, X.~Wu, Y.~Wu, X.~Mao, P.~Xu, and G.~Chen, ``low
  complexity stable link scheduling for maximizing throughput in wireless
  networks,'' in \emph{Proc. IEEE SECON}, 2009, pp. 1--9.

\bibitem{S:constant1}
X.~Lin and S.~B. Rasool, ``Constant-time distributed scheduling policies for ad
  hoc wireless networks,'' in \emph{Proc. IEEE CDC}, 2006, pp. 1258--1263.

\bibitem{S:constant4}
C.~Joo and N.~B. Shroff, ``Performance of random access scheduling schemes in
  multi-hop wireless networks,'' \emph{IEEE/ACM Transactions on Networking},
  vol. $17$, pp. 1481--1493, Oct. 2009.

\bibitem{S:MWM2}
L.~Tassiulas and A.~Ephremides, ``Linear complexity algorithms for maximum
  throughput in radio networks and input queued switches,'' in \emph{Proc. IEEE
  Infocom}, 1998, pp. 533--539.

\bibitem{S:MS}
P.~Chaporkar, K.~Kar, and S.~Sarkar, ``Throughput and fairness guarantees
  through maximal scheduling in wireless networks,'' \emph{IEEE/ACM
  Transactions on Information Theory}, vol. $54$, pp. 572--594, Feb. 2008.

\bibitem{S:phy11}
D.~Chafekar, V.~Kumar, M.~Marathe, S.~Parthasarathy, and A.~Srinivasan,
  ``Arrpoximation algorithms for computing capacity of wireless networks with
  {SINR} constraints,'' in \emph{Proc. IEEE Infocom}, 2008, pp. 1166--1174.

\bibitem{Halldorsson2012ciss}
E.~Asgeirsson, M.~Halld\'{o}rsson, and P.~Mitra, ``A fully distributed
  algorithm for throughput performance in wireless networks,'' in
  \emph{Information Sciences and Systems (CISS)}, 2012, pp. 1 --5.

\bibitem{S:dphy1}
J.~Ryu, C.~Joo, T.~T. Kwon, N.~B. Shroff, and Y.~Choi, ``Distributed {SINR}
  based scheduling algorithm for multi-hop wireless networks,'' in \emph{Proc.
  ACM MSWIM}, 2010, pp. 376--380.

\bibitem{modiano2009wiopt}
H.-W. Lee, E.~Modiano, and L.~B. Le, ``Distributed throughput maximization in
  wireless networks via random power allocation,'' in \emph{Modeling and
  Optimization in Mobile, Ad Hoc, and Wireless Networks (WiOPT)}, 2009, pp. 1
  --9.

\bibitem{Halldorsson2011soda}
M.~M. Halld\'{o}rsson and P.~Mitra, ``Wireless capacity with oblivious power in
  general metrics,'' in \emph{Proceedings of the Twenty-Second Annual ACM-SIAM
  Symposium on Discrete Algorithms}, 2011, pp. 1538--1548.

\bibitem{Akummar2012}
P.~Guanhong and V.~S.~A. Kumar, \emph{Distributed Link Scheduling under the
  Physical Interference Model}, 2012, manuscript.

\bibitem{S:phy17}
D.~M. blough, G.~Resta, and P.~Santi, ``Approximation algorithms for wireless
  link scheduling with {SINR-Based} inteference,'' \emph{IEEE/ACM Transactions
  on Networking}, vol. $18$, pp. 1701--1712, Dec. 2010.

\end{thebibliography}

\appendix
\subsection{Proof of Lemma 1}

\begin{IEEEproof}
Our proof bases on the fact of fading metrics \cite{halldorsson_2009}. In fading metrics the path loss exponent $\kappa$ must be strictly greater than the doubling dimension of the metric, and the doubling dimension $A=n$ for the $n-$dimensional Euclidean space.
We have assumed the Euclidean plane and the path loss exponent $\kappa > 2$, obviously these assumptions construct a fading metric of doubling dimension $A=2$. For the fading metric of doubling dimension $A$,  there are at most $Cg^A$ balls of radius $Z$ inside a ball of radius $gZ$ for any $g > 0$. Here $C=\frac{1}{6}\pi \sqrt{3} \approx 0.907$ for the Euclidean plane. A ball of radius $\mu$, centered at $v$ is defined by $B(v,\mu)$.

Let{\small{ $X_g = \{w \in V(L) | d(w,v) < gd \slash 2 \}$ }}for $g>0$.
The distance between any two nodes in $V(L)$ is at least $d$. It  implies $B(v,(g+1)d \slash 2)$ contains all balls of radius of $d \slash 2$ centered at the nodes in $X_g$ and these balls do not intersect. It is obvious that $|X_2|=0$ for the smallest mutual distance between any pair of nodes is $d$. Then for each node $v \in V(L), $ it holds that,
{\small{
\begin{eqnarray*}
&~&     \sum \limits_{w \in V(L)} {\frac{R^{\kappa}}{d(w,v)^{\kappa}}}                            \\*
&\leq&   \sum_{g=3}^{\infty}|X_{g} \backslash X_{g-1}| \frac{R^{\kappa}}{[(g-1) d \slash 2]^{\kappa}} \\*
&\leq& \frac{R^{\kappa}}{(d \slash 2)^{\kappa}} \cdot \sum_{g=3}^{\infty} |X_{g}| \left(\frac{1}{(g-1)^{\kappa}}-\frac{1}{g^{\kappa}}\right)                                                \\*
&\leq& \frac{R^{\kappa}}{(d \slash 2)^{\kappa}} \cdot \sum_{g=3}^{\infty} |X_{g}| \frac{\kappa}{(g-1)^{\kappa+1}} \\*
&\leq& \frac{R^{\kappa}}{(d \slash 2)^{\kappa}} \sum_{g=3}^{\infty} C \cdot (g+1)^{A} \frac{\kappa}{(g-1)^{\kappa+1}}                                                                            \\*
&\leq&  \frac{R^{\kappa}}{(d \slash 2)^{\kappa}} \sum_{g=3}^{\infty} C \cdot  \frac{\kappa (g+1)^{A} 2^{\kappa+1}}{(g+1)^{\kappa+1}}                                                                       \\*
&<& \frac{ 2^{2\kappa+1}\kappa C}{\theta^{\kappa} (\kappa-A)}   \\*
&=& \frac{ 2^{2\kappa+1} \sqrt{3} \pi   \kappa }{6 (\kappa-2) \theta^{\kappa} } {~} ={~} O(1/\theta^{\kappa})
\end{eqnarray*}
}}
\end{IEEEproof}

\end{document}